\documentclass[useAMS,usenatbib]{mnras}
\usepackage{epsfig,appendix,array,rotating,pdflscape,array,longtable,colortbl,appendix,footnote,cancel}
\usepackage{multirow,url}
\usepackage{booktabs}
\usepackage{lscape, amsmath} 
\usepackage{paralist}
\usepackage{relsize}
\usepackage{amssymb}
\usepackage{xtab}
\usepackage{hyperref}
\usepackage[normalem]{ulem}
\DeclareUnicodeCharacter{2212}{-}

\newcommand{\hi}{H\,{\sc i}}

\newcommand{\prim}{$^{\prime}$}
\newcommand{\prin}{$^{\prime\prime}$}

\newcommand{\km}{km\,s$^{-1}$}
\newcommand{\degree}{$^{\circ}$}
\newcommand{\halpha}{H${\alpha}$}
\newcommand{\cg}{CGCG\,097-}
\newcommand{\msolar}{M$_{\odot}$}

\newcommand{\mhi}{$M({\mathrm {HI}})$}

\newcommand{\defhi}{H\,{\sc i} deficiency}

\newcommand{\vrel}{\textit{v$_\mathrm{\textit{rel}}$}}

\newcommand{\angs}{${\mathrm {\AA}}$}
\newcommand{\msolaryr}{M$_{\odot}$\,yr$^{-1}$}

\newcommand {\apgt} {\ {\raise-.5ex\hbox{$\buildrel>\over\sim$}}\ }
\newcommand {\aplt} {\ {\raise-.5ex\hbox{$\buildrel<\over\sim$}}\ }

\newcommand{\af}{$A_\mathit{flux}$}
\setlongtables

\newcommand\xoutpars[1]{\let\helpcmd\xout\parhelp#1\par\relax\relax}
\newcommand\soutpars[1]{\let\helpcmd\sout\parhelp#1\par\relax\relax}
\long\def\parhelp#1\par#2\relax{%
  \helpcmd{#1}\ifx\relax#2\else\par\parhelp#2\relax\fi%
}

\title[FGC\,1287's 250\,kpc long HI tail]{FGC\,1287 and its enigmatic 250\,kpc long HI tail in the outskirts of Abell\,1367}
\author[]{T.\ C.\ Scott$^{1}$\thanks{E-mail: tom.scott@astro.up.pt (TCS)}, L.\ Cortese$^{2,3}$, P.\ Lagos$^{1}$, E.\ Brinks$^{4}$,  A.\ Finoguenov$^{5}$ and L.\ Coccato $^{6}$\\
$^{1}$Institute of Astrophysics and Space Sciences (IA), Rua das Estrelas, 4150--762 Porto, Portugal\\
$^{2}$International Centre for Radio Astronomy Research, The University of Western Australia, 7 Fairway, 6009, Crawley WA, Australia\\
$^{3}$\textcolor{black}{ARC Centre of Excellence for All Sky Astrophysics in 3 Dimensions (ASTRO 3D), Australia}\\
$^{4}$Centre for Astrophysics Research, University of Hertfordshire, College Lane, Hatfield, AL10 9AB, UK \\
$^{5}$University of Helsinki, Department of Physics, Gustaf Hallstromin katu 2 00140 Helsinki, Finland\\
$^{6}$European Southern Observatory, Karl-Schwarzchild-str., 2, 85748 Garching,  Germany\\
}

\begin{document}
\date{Accepted. Received ; in original form }


\maketitle

\label{firstpage}

\begin{abstract}
We present \hi\ and radio continuum, narrow--band H$\alpha$ imaging,  IFU spectroscopy, and X--ray observations of the FGC\,1287 triplet projected $\sim$ 1.8 Mpc west of the galaxy cluster Abell\,1367.  One triplet member,  FGC\,1287, displays an exceptionally long, 250\,kpc \hi\  tail and an unperturbed stellar disk which are the typical  signatures of ram pressure stripping (RPS). To generate detectable RPS signatures the presence of an Intra--cluster medium ICM/intra--group medium IGM with sufficient density to produce RPS at a realistic velocity relative to the ICM\textcolor{black}{/IGM} is a prerequisite. \textcolor{black}{However, {\em XMM--Newton} observations were not able to detect X--ray emission from the triplet, implying that if a hot ICM/IGM is present, its density, }  $n_e$, is less than 2.6$ \times 10^{-5}$\,cm$^{-3}$.   Higher--resolution VLA \hi\ data presented here  show  FGC\,1287's \hi\ disk is truncated and significantly warped whereas the  \hi\ tail is clumpy. TNG H$\alpha$ imaging identified three star forming clumps projected within 20 kpc of  FGC\,1287's disk, with VIMOS--IFU data confirming two of these are counterparts to \hi\ clumps in the tail. The triplet's \hi\ kinematics, together with H$\alpha$ and radio continuum imaging suggests an interaction  may have enhanced star formation in FGC\,1287's  disk, but cannot readily account for the origin of the long \hi\ tail. We consider several scenarios which might reconcile RPS with the non--detection of ICM/IGM X--ray emission but none of these unambiguously explains the origin of the long \hi\ tail. 

\end{abstract}

\begin{keywords} galaxies:clusters:individual: (Abell\,1367) galaxies: groups: individual: FGC\,1287 triplet  galaxies: ISM 
\end{keywords}

\section{INTRODUCTION}
\label{into}
Details of how, when and where the galaxy cluster environment impacts late--type galaxies (LTGs) infalling into clusters remains to be definitively determined.  \hi\ observations from low-{\em z} clusters ($z \lesssim 0.03$) suggest that hydrodynamic interactions with the intra--cluster medium (ICM), e.g., ram pressure stripping, viscous stripping, thermal evaporation  and starvation  are  the dominant mechanisms removing the interstellar medium (ISM) from cluster LTGs and accelerating their evolution toward earlier morphological types \citep{bosel06a,cort21}. Hydrodynamic stripping of the ISM (primarily \hi) from  cluster LTGs is caused by their interaction with  the ICM during their orbits through a cluster.   Within a few hundred kpc of cluster cores, ICM electron number densities ($n_e$) are $\geq$ 10$^{-3}$  cm$^{-3}$  \citep{bacall99} and galaxy velocities relative to the ICM (\vrel) can exceed  $\sim$ 1000 \km. Under these conditions ram pressure stripping, thermal evaporation and viscous stripping, collectively ``RPS",  are a viable mechanism to explain most of the observed ongoing cold and ionized gas stripping signatures on time scales of a few $\times$ 10$^8$ yr  \citep[e.g.,][]{gava95,bravo00,voll04,chung09,yagi17,jachym19,poggianti19}. RPS  could  also feasibly account for most of the observed \hi\ deficiencies and star formation (SF) quenching in LTGs within the virial radii of nearby clusters \citep{sola01,cort12}. The ram pressure ($P_{ram}$)  a cluster galaxy is subject to depends on a combination of ICM density ($\rho_{ICM}$) and its velocity relative to the ICM (\vrel), i.e., $P_{ram}$ = $\rho_{ICM}$ \vrel$^2$ \citep{gun72}.    

\begin{figure}
\begin{center}
\includegraphics[ angle=0,scale=0.6] {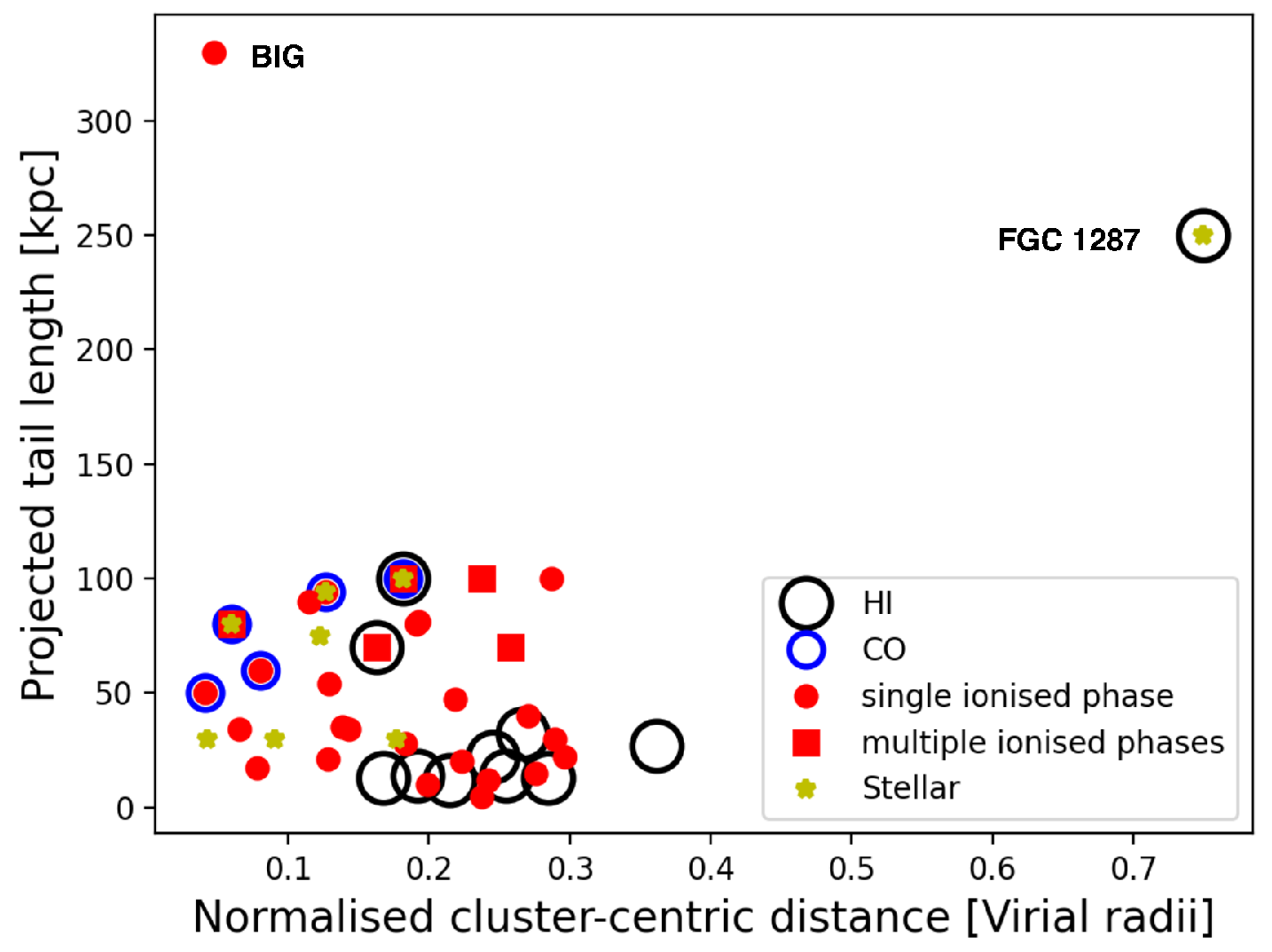}
\vspace{0.01cm}
\caption{Cluster galaxies with one--sided tails \textcolor{black}{taken from a literature sample of galaxies with one--sided tails as well as  FGC\,1287 and the BIG group (see main text for details)}. Plotted is projected  tail length against distance from the \textcolor{black}{cluster centre}  normalised by \textcolor{black}{cluster} virial radius. The symbols indicate the different types of tails: \textcolor{black}{circles = cold gas (\hi\ -- black, CO -- blue), red = single or a combination of tracers of ionised gas (\halpha,UV, X--ray) and/or radio continuum), i.e. plasma tails and stars = stellar.}}
\label{fig_1}
\end{center}
\end{figure}

\textcolor{black}{One--sided} tails of galaxies in clusters, consisting variously of combinations of stars, cold  and ionised gas, have commonly been attributed to RPS.   Figure \ref{fig_1}  shows the projected tail length of a representative sample of 44  galaxies with one--sided RPS tails from 9 clusters, as a function of their distance from the cluster centre and normalised by cluster virial radius. \textcolor{black}{Where a galaxy has a  multi--phase tail, the tail length in the plot is for the phase with the largest projected length.} The  data for the galaxies  shown in the plot is from papers by \citet{gava87,gava95,gava01b,cort07,chung07,yoshida08,hester10,sun10, scott10,scott18,smith10,yagi10,yagi17,bosel16,jachym17,jachym19,poggiant17,poggianti19, bellhouse19,
Ramatsoku19,moretti20,muller21}. The plot includes the tails observed for the Blue Infalling Group (BIG) and FGC\,1287, both in Abell\,1367. All of the RPS tails, except FGC\,1287, are projected within $\sim$ 40\% of their cluster's virial radius ($\approx$ 1 Mpc) and are  predominantly found in large, ICM rich clusters. In several cases, RPS appears to be aided by tidal interactions and/or AGN activity  "loosening"  gas from the parent galaxy's gravitational potential well  \citep[e.g.,][]{cosolandi17,poggianti20}.  FGC\,1287's uniqueness is clear from this plot both in terms of its distance from the cluster centre and the length of its tail.

Beyond the  virial radii of clusters,  it is less clear  whether RPS significantly impacts LTGs. This is primarily because  X--ray emission from the ICM has not been detected there and, while we expect that low density warm medium undetectable by X--ray telescopes should be present \citep{cen01,macquart20}, it is still debated whether hydrodynamical effects remain the dominant environmental mechanism affecting galaxy evolution in the outskirts. Indeed, many authors, e.g., \cite{dress04},  \textcolor{black}{argue} that in these environments prolonged exposure to intra--group tidal interactions and ``preprocessing" more generally, is more efficient in affecting the cold gas reservoirs of galaxies. Evidence of the impact of preprocessing on the \hi\ in LTGs in a rich infalling group was recently presented in \citet[][]{kleiner21}, \textcolor{black}{see also} \cite{cort21}.

To settle this debate, further resolved \hi\ observations of galaxy  clusters are required as the \hi\  morphology and kinematics of disturbed galaxies can provide important insights into the mechanisms at play. In this paper we extend our previous analysis of one of the most dramatic examples of \textcolor{black}{a one--sided}  \hi\ tail in the cluster outskirts: FGC\,1287 \citep[][]{scott12}. Despite being at a projected distance $\sim$ 1.8\,Mpc west of the centre of A\,1367, this galaxy shows an extraordinarily \textcolor{black}{long}  250\,kpc \hi\ tail, significantly longer than the typical HI tails associated with RPS galaxies near  the centres of clusters.

In this paper we report on high resolution NRAO\footnote{The National Radio Astronomy Observatory is a facility of the National Science Foundation operated under cooperative agreement by Associated Universities, Inc.} Karl G. Jansky Very Large Array (VLA) B-- and C--configuration \hi\ and radio continuum,  ESO\footnote{European Southern Observatory} VLT/VIMOS\footnote{Very Large Telescope (VLT)}, VIsible Multi--Object Spectrograph (VIMOS) integral field unit (IFU) spectroscopy (hereafter VIMOS--IFU) and {\em XMM--Newton} X--ray observations of the FGC\,1287 triplet. Additionally, we present narrow band  \halpha\ imaging of FGC\,1287 from the Telescopio Nazionale Galileo (TNG) 3.58-m telescope located at Roque de Los Muchachos, La Palma, Spain.
 
Section \ref{obs} summarises the    observations,  with  observational results in section \ref{results}. A discussion follows in section \ref{discussion} with concluding comments in section \ref{concl}.  Based on a redshift to A\,1367 of 0.022 and assuming $\Omega_M =0.3$, $\Omega_\Lambda =0.7$, and  $H_o  =72$\,\km\,Mpc$^{-1}$ \citep{sperg07} the distance to the cluster is $\sim$ 92\,Mpc and the   angular scale is 1\,arcmin \textcolor{black}{=} 24.8\,kpc. All projected positions referred to throughout this paper are in J2000.0. 

\section{OBSERVATIONS}
\label{obs} 
\subsection{VLA}
\label{obs_vla}
The FGC\,1287 triplet was observed with  the VLA in B-- (9.2 hrs on source) and C-- configurations (3.2 hrs)  in L--band  (1420\,MHz; project ID:13B--019). A summary of each day's observation is set out in Table \ref{table1}.  

Each  day's observation was configured to record 10 spectral windows (spw), 8 of which contained  64 continuum channels each at full polarisation and with the other  two spws set up with 1028 channels for spectral line observations, in dual polarisation mode.  Spw 3 was set up to observe the \hi\ line of neutral atomic hydrogen and generated a position--position--velocity cube ($\alpha,  \delta, velocity $) with  1028 (7.81\,kHz wide) channels. The remaining spw was centred on the OH transitions which were, however, not detected \textcolor{black}{and no further use was made of this spw}.

Each day's measurement set  was reduced with the VLA pipeline provided within the Common Astronomy Software Applications package {\sc (casa)} following standard procedures \citep{mcmull07}.  After verifying each day's observations reached the required quality and expected noise value,  we combined the visibilities of the \hi\ measurement sets (spw 3)   into a single combined (B/C configuration)  measurement set using the {\sc casa} task {\sc concat}.  We removed the continuum from this combined measurement set using the task {\sc uvsub}. To further reduce the effect of RFI  in this cube we inspected the continuum subtracted {\em uv}--data and  flagged visibilities with amplitudes $>$ 2.5\,Jy.  On this \hi\ continuum subtracted, combined  B/C configuration measurement set we used the task {\sc tclean} and produced a cube with with 352 channels  (barycentric optical velocity range of 5953\,\km\ to 7708\,\km) with a  reduced velocity resolution of 5\,\km,  to increase the S/N (rms per channel of $\sim$ 0.32\,mJy). The synthesised beam for this cube was  13.84 arcsec  $\times$ 12.69 arcsec,  PA=178.09\degree. Because we were particularly interested in the faint \hi\ emission in the FGC\,1287 tail we produced the B/C configuration \hi\ cube  using natural weights. Natural weighting improves sensitivity at the cost of angular resolution. \textcolor{black}{No self--calibration was required.} 

For the B/C configuration \hi\ cube the \textcolor{black}{3$\sigma$} detection limit \textcolor{black}{ in a three channel average } was N(HI) = \textcolor{black}{5} $\times$ 10$^{19}$\,atom\,cm$^{-2}$ (2  $\times$ 10$^7$\,\msolar). \textcolor{black}{A three channel average corresponds to a linewidth of ~15 km\,s$^{-1}$ which in turn corresponds to a velocity dispersion of 6 \km\ which is what is typically found in the outskirts of disk galaxies \citep{tamburu09}. \hi\ moment 0 and 1 maps \textcolor{black}{(integrated \hi\ and velocity field, respectively)}  were extracted from the final  B/C configuration cube (velocity resolution of 5 \km) after blanking carried out in the following manner with {\sc casa}:
1) the \hi\ cube was spatially smoothed with a boxcar 3 $\times$ 3 filter.
2) pixels in all channels beyond the previous D--configuration moment 0 3$\sigma$ contour were masked (i.e., discarded). 
3) all remaining pixels with fluxes below a 2$\sigma$ clip were masked
4) channels with velocities where no AGES \hi\ flux was detected in the triplet or AGES J113939+193524 were fully blanked.
5) only those pixels were retained which formed part of coherent structures in velocity by comparing for each channel what emission was present in adjacent channels. This was done manually and had the effect of removing noise peaks that are above the 2$\sigma$ clip threshold in one channel only.
6) the masks from the previous steps were combined and applied to the original unsmoothed B/C configuration cube to produce a blanked cube to which the {\sc casa} task  {\sc imoments} was applied to extract the moment 0 and 1 \hi\ maps. 
} 
For the 8 continuum spws we carried out additional flagging of the calibrated B and C configuration measurement sets using AOFlagger. 





\begin{table*}
\footnotesize
\begin{minipage}{150mm}
\caption{VLA observational parameters}
\label{table1}
\begin{tabular}{@{}cccccrrr@{}}
\hline

Date&Configuration&
Pointing & centre & Int.\footnote{On target integration time.}& Central\footnote{Barycentric optical velocity of the \hi\ spectral window. }   \\
&& ${\alpha}_{2000}$ & ${\delta}_{2000}$ &time &Velocity   \\
&& [$^h$ $^m$ $^s$] & [\degree\ \prim\ \prin\ ] &  [hours]& [\km ]  \\ \hline 
14 Nov 2013 &B&  11 39 08.3 & 19 39 02.2 & 2.3&  6830  \\
15 Nov 2013 &B&  11 39 08.3 & 19 39 02.2 & 2.3&  6830  \\
24 Nov 2013 &B&  11 39 08.3 & 19 39 02.2 & 2.3&  6830  \\
30 Nov 2013 &B&  11 39 08.3 & 19 39 02.2 & 2.3&  6830  \\
\,\,6  Dec 2014 &C & 11 39 08.3 & 19 39 02.2& 3.2& 6830 \\
 \hline
\end{tabular}
\end{minipage}
\end{table*}

\subsection{VIMOS--IFU}

 \textcolor{black}{The observations were obtained using VIMOS--IFU on the 8.2 m VLT UT3/Melipal telescope in Chile. The VIMOS--IFU consists of four CCD quadrants covered by a pattern of 1600 elements. We used a projected size per element of 0.33", covering a total field of view of 13" $\times$ 13". Four VIMOS--IFU pointings were obtained on the 10th and 16th of February 2016, two at the optical centre of FGC\,1287 and two at the positions of \halpha\ emitting clumps (SFC1 and SFC2) north of the optical centre (see Figure \ref{fig_2}). Two science exposures of 830 s were taken per pointing. A third dithered exposure of 290 s was taken, after each observation, in order to obtain a night sky background exposure.These observations used the High Resolution (HR) HR--orange grating (0.62 \angs\ pixel$^{-1}$), offering a spectral resolving power R = 2650 between $\sim$ 5100--7699 \angs. The data were reduced using the \textit{esorex} software version 3.10.2. This included bias subtraction, flat--field correction, wavelength and flux calibration. The flux calibration was based on observations of spectrophotometric standard stars included in the standard VIMOS calibration plan. The 2D data images were combined into 3D data cubes and re-sampled to a 0.33" spatial resolution. Finally, we correct the resulting data cubes for the quadrant--to--quadrant intensity differences. We renormalized the quadrants by comparing the intensity levels of the neighbouring pixels at the quadrant borders. More details about the data reduction can be found in \cite{lagos16,lagos18}.} 

\subsection{ \halpha\ imaging} 

FGC\,1287 was imaged with the Device Optimized for  LOw RESolution (DOLORES) attached at the Nasmyth B focus of the 3.6-m Telescopio National Galileo on 23rd February and 15th March, 2012. Narrow--band imaging was obtained using a [SII] narrow band filter centred at $\sim$6724 \AA\ with a width of $\sim$57 \AA\ to include the red--shifted H$\alpha$ line, with the underlying continuum estimated via broadband (Gunn) r' filter observations. Total exposure time was 9600 sec and 7560 sec for the narrow and broad--band observations, respectively. Data reduction was performed using standard techniques in IRAF, following the same procedure as described in \cite{cort06}.\textcolor{black}{That is all single science frames were stacked after individual bias subtraction and flat-fielding, using the {\sc incobine} and {\sc imarit} tasks. The combined OFF--band frame was then normalised to the combined ON--band one by the flux ratio of a few foreground stars. The H$\alpha$+[NII] NET image shown in the lower right panel of Fig. \ref{fig_2} was then obtained by subtracting the normalized OFF--band image from the ON--band one.}

\subsection{XMM--Newton} 

{\em XMM--Newton} observed  FGC\,1287 on 16 June 2012 for a duration of 18.4 ksec, observational ID 0690780101. The thin filter was used with all EPIC cameras, with  total clean exposures of 11.0, 17.6 and 16.1 ksec for PN, MOS1 and MOS2, respectively. The observational setup used is sensitive to the X--ray emission of the gas with temperatures above ~0.1 keV. Lower temperature gas in collisional ionization equilibrium would remain undetected in this experiment.

The source was not detected in the combined 0.5--2 keV band image. Using a 5 arcmin  radius centred on FGC\,1287, we extracted PN spectrum and adopted a 2$\sigma$ limit to normalise the APEC plasma model using a temperature of 0.5 keV, leading to a corresponding limit on the gas mass of 7 $\times$ 10$^9$\msolar, assuming a uniform gas distribution.


\section{OBSERVATIONAL RESULTS}
\label{results}
\subsection{HI and Radio continuum properties}
\label{results_hi}
\begin{figure*}
\begin{center}
\includegraphics[ angle=0,scale=0.84] {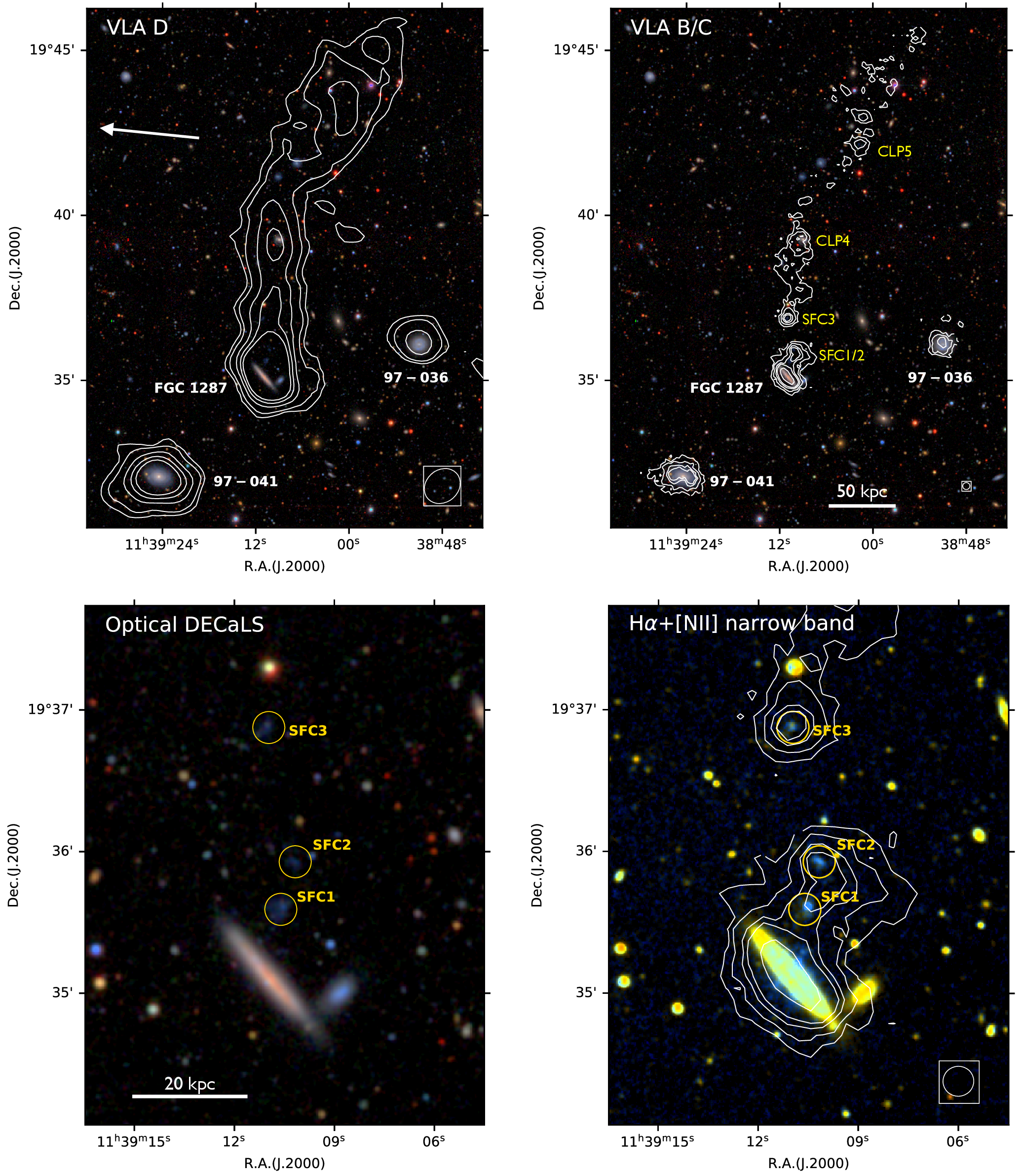}
\caption{FGC\,1287 group: (\textit{top left:)} \hi\  VLA D--configuration \hi\ \textcolor{black}{column} density map as contours on an SDSS $g,r,i$--band image, coded as BGR \citep{scott12}.  \textcolor{black}{Contours are at 1.6, 2.6, 5.2, 7.8 and 10.3 $\times$ 10$^{19}$ atoms cm$^{-2}$.}  The boxed ellipse in the lower right shows the  size  ($72^{\prime\prime} \times 60^{\prime\prime}$)  and orientation of the VLA D--configuration synthesised beam. The arrow points in the direction of the A\,1367 cluster centre. (\textit{top right:)} \hi\  VLA B/C--configuration \hi\ \textcolor{black}{column} density map as contours on an SDSS $g,r,i$--band image, coded as BGR.  \hi\ contours are at column densities of 0.5, 1.6, 3.1, 4.7, 6.2, 9.3, and $12.5 \times 10^{20}$\,atom\,cm$^{-2}$. The three \hi\ detected galaxies are marked together with the projected positions of principal \hi\ clumps in the FGC\,1287 tail.  (\textit{bottom left:}) Optical  \textcolor{black}{$g,r,z$ DECaLS} image of the region at the base of the FGC\,1287 \hi\ tail.  The small blue galaxy just above the SW edge of  FGC\,1287 is ASK 626970 which is projected in the foreground of the triplet. (\textit{bottom right:}) TNG \halpha+[NII] narrow band image (blue) of the  the region at the base of the FGC\,1287 \hi\ tail overlaid with VLA B/C--configuration \hi\ contours (contour levels as per top right panel.  The beam size and orientation of VLA B/C--configuration synthesised beam is indicated by the boxed ellipse at the bottom right and measures $13.8^{\prime\prime}  \times 12.7^{\prime\prime}$).} 
\label{fig_2}
\end{center}
\end{figure*}

The VLA B/C--configuration  \hi\ \textcolor{black}{column} density (henceforth referred to as moment 0) maps for FGC\,1287 and neighbouring galaxies \cg036 and \cg041 reveal an \hi\ tail clumpier than originally suggested by the lower resolution \textcolor{black}{VLA---D configuration} observations presented in \cite{scott12}.  Figure \textcolor{black}{ \ref{fig_2} shows} at least five  separate structures above an \hi\ column density of $\sim$3.1  $\times$ 10$^{20}$ atoms cm$^{-2}$ are visible within the \hi\ tail (labeled  SFC1, SFC2, SFC3, CLP4 and CLP5).  \hi\ was also detected in AGES J113939+193524   \citep[][]{cort08}, \textcolor{black}{which is} the \hi\ counterpart of the optical dwarf galaxy SDSS J113939.67+193516.7, see  Appendix A.    Properties of the VLA \hi\ detections are summarised in \textcolor{black}{Tables \ref{table2} and \ref{table2a}.} \textcolor{black}{As a check we also applied the \textsc{SOFIA--2}  \citep{serra15}  source finding code to the \textcolor{black}{unmasked} B/C configuration \hi\ cube which generated moment maps in good agreement with those produced with our manual blanking  procedure \textcolor{black}{(Section \ref{obs_vla})}, recovering all five \hi\ tail clumps with $\geq$ 90\% reliability. } 

\begin{figure*}
\begin{center}
\includegraphics[ angle=0,scale=0.8] {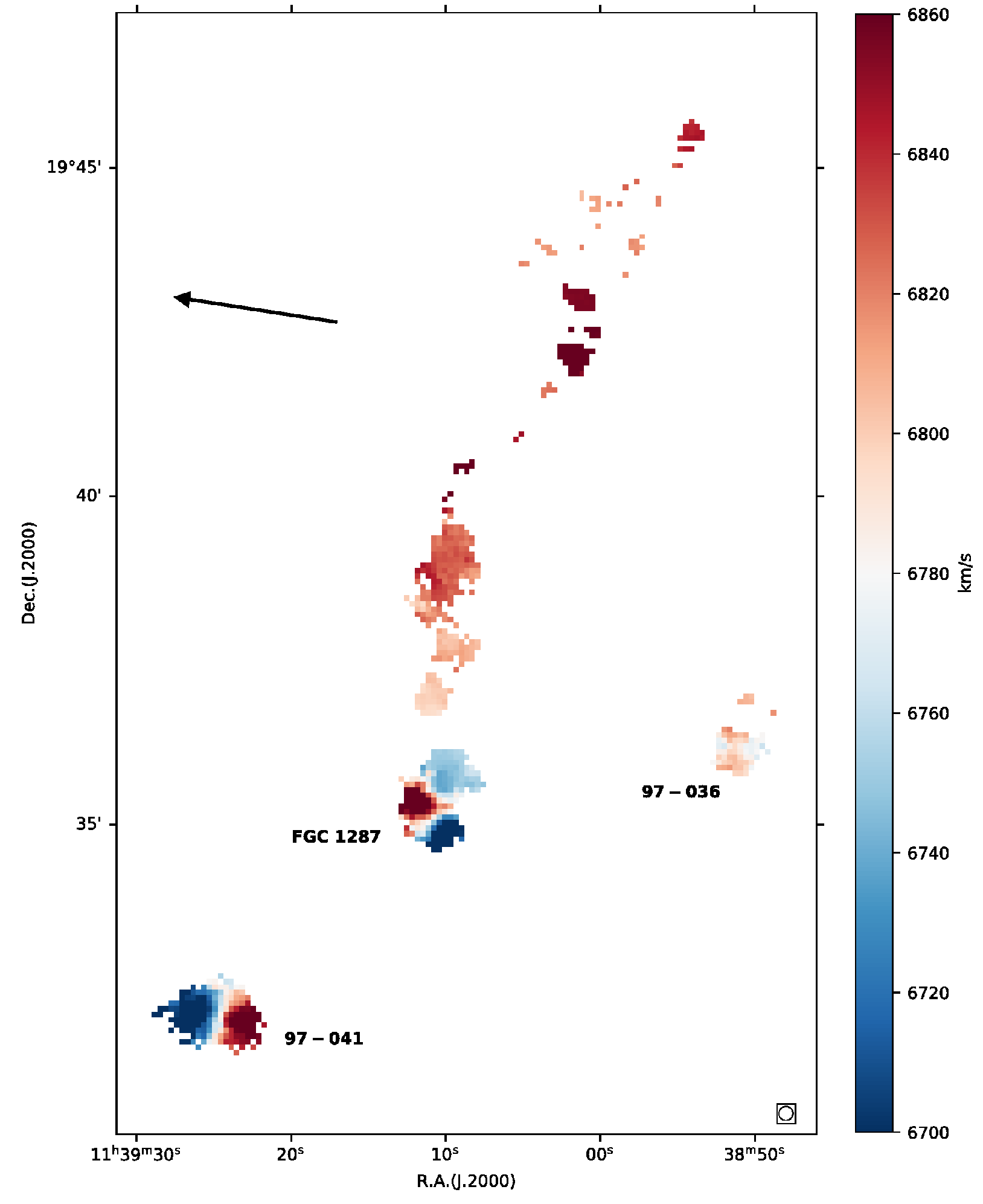}
\caption{FGC\,1287 group: \hi\ velocity field\textcolor{black}{, with velocities corresponding to the colours in the sidebar. }  The beam size and orientation of VLA B/C--configuration synthesised beam is indicated by the boxed ellipse in the bottom right and measures $13.8^{\prime\prime}  \times 12.7^{\prime\prime}$. The arrow points in the direction of the cluster centre of A\,1367. }  
\label{fig_2a}
\end{center}
\end{figure*}

\begin{figure*}
\begin{center}
\includegraphics[ angle=0,scale=0.83] {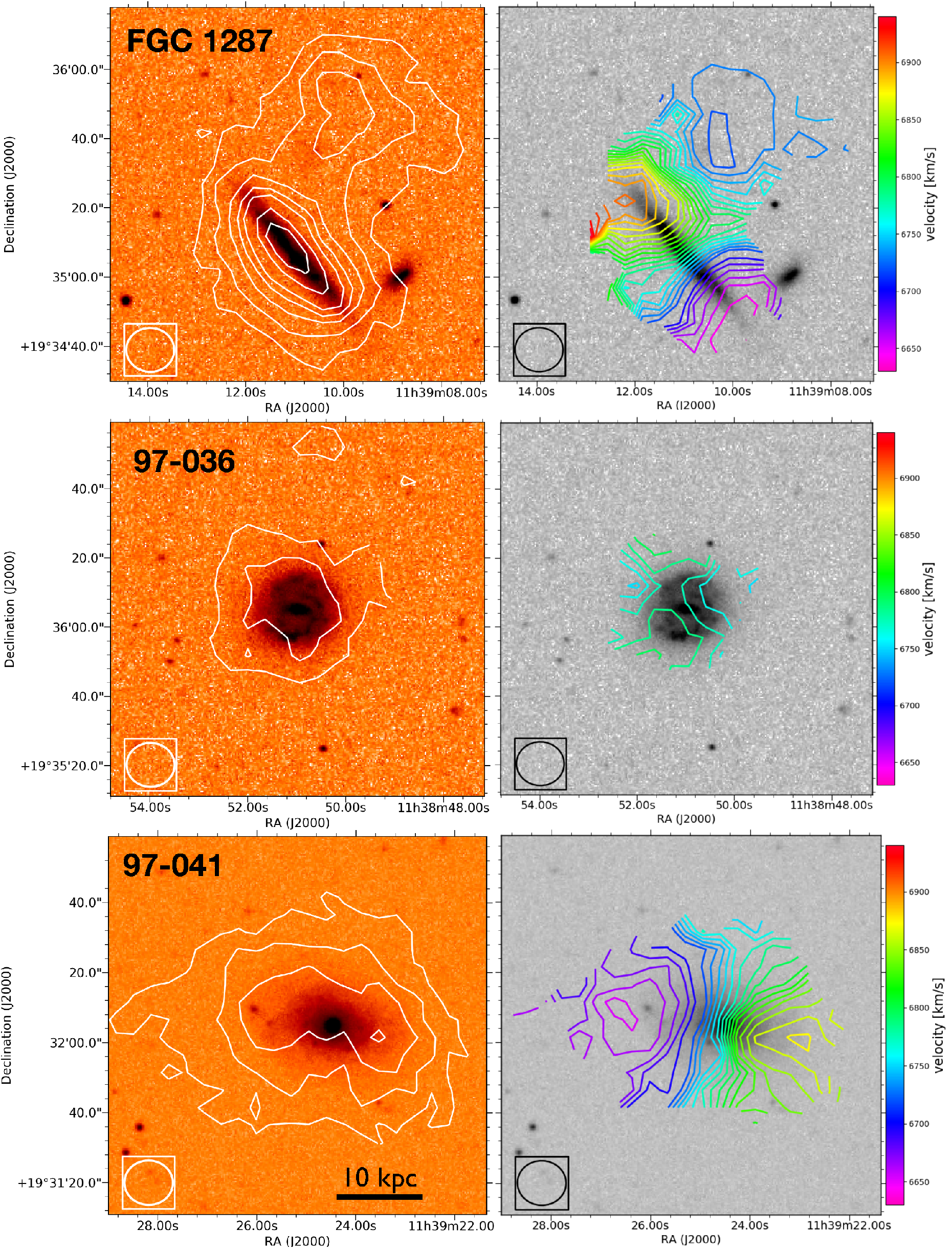}
\vspace{1cm}
\caption{Galaxies in the FGC\,1287 group: (\textit{\textbf{left column}}) \hi\ \textcolor{black}{column} density as white contours from the VLA B/C-configuration for  FGC\,1287, \cg036, and \cg41. The contours are at \hi\ column densities of 0.5, 1.6, 3.1, 4.7, 6.2, 9.3, and 12.5 $\times$ 10$^{20}$\,atom\,cm$^{-2}$, i.e., the same as in Figure \ref{fig_2}. (\textit{\textbf{right column}})  \hi\ velocity fields with contours separated by 10 \km, except  for \cg36 where the separation is 20 \km.  The background images are from the SDSS $g$--band. The beam size and orientation of VLA B/C--configuration synthesised beam is indicated by the boxed ellipse in the bottom right and measures $13.8^{\prime\prime}  \times 12.7^{\prime\prime}$). }
\label{fig_3}
\end{center}
\end{figure*}

\begin{figure}
\begin{center}
\includegraphics[ angle=0,scale=.48] {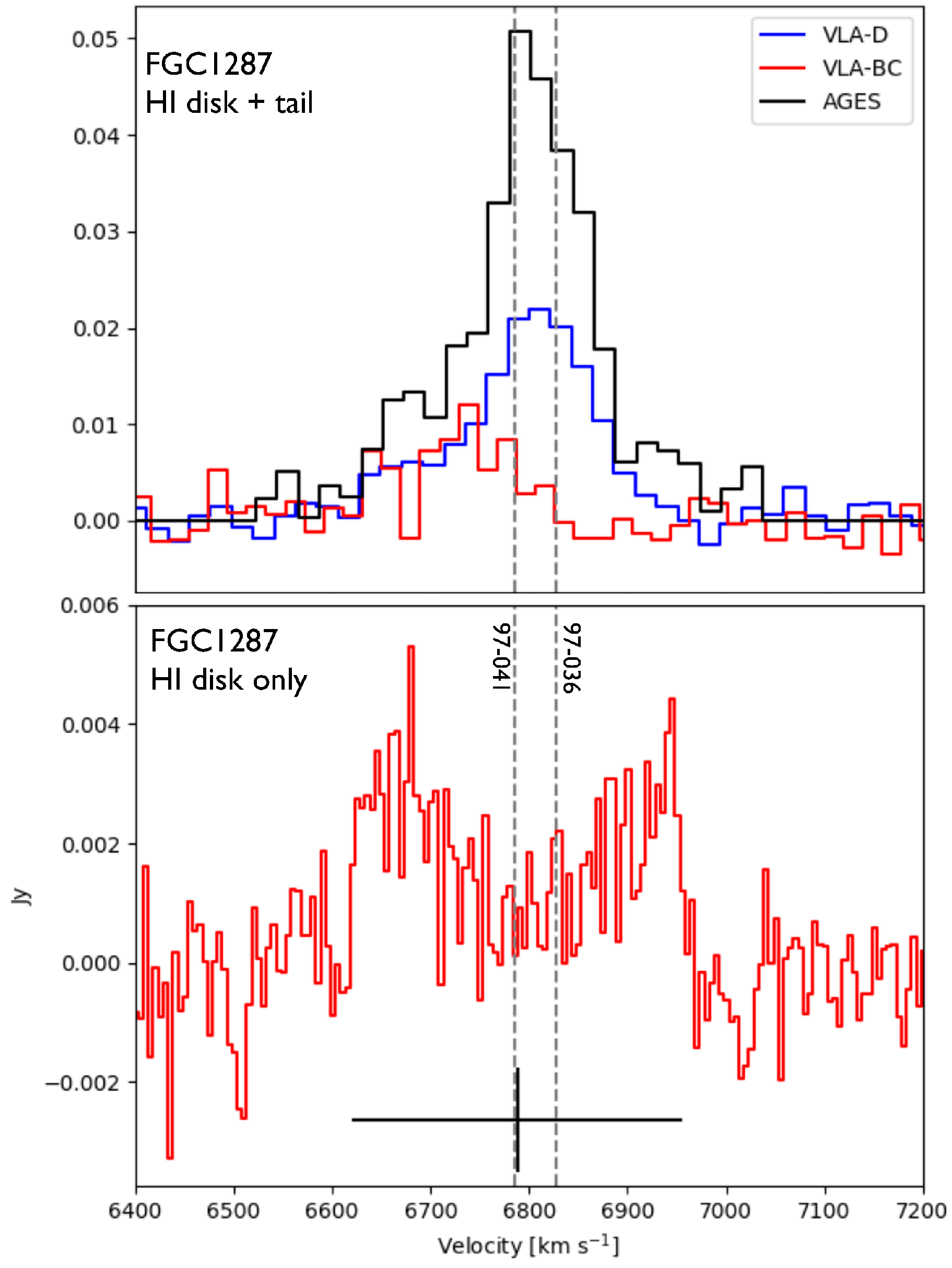}
\vspace{1cm}
\caption{\textcolor{black}{\textbf{FGC\,1287 \textit{top panel:} } comparison between the \hi\ spectra for the FGC\,1287 disk + tail from the  AGES as well as the VLA D-- and VLA B/C--configurations per the legend. The channel width of each  spectrum is 21\,\km. The spectra show the substantial flux loss due to the lower column density sensitivity of the VLA observations and lack of short spacings.   \textbf{\textit{Lower panel:} } \hi\ spectrum for the FGC\,1287 disk component extracted from the  VLA B/C--configuration cube. The cross at the base of that spectrum indicates the  disk component's \hi\ W$_{20}$ and V$_{HI}$.} Vertical dashed lines indicate the optical velocities of \cg036 and \cg041. }
\label{fig_4}
\end{center}
\end{figure}

The velocity field for the FGC\,1287 group is shown in Figure \ref{fig_2a}, with a zoom--in  of the \hi\ \textcolor{black}{moment 0} map and  velocity field for each group member shown in Figure \ref{fig_3}.  Whereas  the SW side of the FGC\,1287 disk shows a regular rotation pattern in the velocity field, the NE part reveals a rapidly changing position angle (PA), between $\sim$ 6850 \km\ to 6900 \km, and closed iso--velocity contours indicating a serious warp. This feature is  also seen in the  D--configuration \hi\ velocity field. \textcolor{black}{The lower panel of Figure \ref{fig_4} presents the spatially integrated \hi\  profile for the FGC\,1287 disk extracted from the VLA B/C--configuration cube. This displays the double horn  profile expected from an edge--on  rotating disk. \textcolor{black}{Kinematic modelling of the \hi\ disk, e.g., with \textsc{$^{3{\mathrm D}}$barolo}  \citep{DiTeodoro15}, was not carried out because the small number of beams across the FGC\,1287 \hi\ disk and the evidence of an asymmetric warp from the \hi\ velocity field made it clear that any model fit obtained would be highly uncertain. } Another kinematic feature is a sequential increase in the velocity of the principal \hi\ clumps in the tail with increasing distance from optical disk, see Table \ref{table3}. The clump velocities  fall within $\pm$ 75 \km\ of the systemic velocity of the FGC\,1287 disk, i.e., the radial velocities of the tail clumps are closer  to the V$_{HI}$ of  FGC\,1287 and the  other two triplet members  than velocities  at the edge of the \hi\ disk. The dashed vertical lines in Figure \ref{fig_4} correspond to the optical velocities of \cg036 and \cg041}.

\textcolor{black}{In the upper panel of Figure \ref{fig_4}   we compare  the integrated \hi\  profiles for the FGC\,1287 disk + tail from the VLA B/C--  and VLA D--configurations as well as from a new extended AGES survey (Deshev et al. 2022 in prep.), referred from here on as AGES. The VLA--D configuration has a 3$\sigma$ \hi\ column density detection limit of $\sim$ 2 $\times$ 10$^{19}$ atoms cm$^{-2}$ \citep{scott12}. From the figure it is clear that  the VLA configurations, and particular the B/C configuration, are missing substantial fractions  of  \hi\ flux. In the case of the B/C array this is due to a combination of a lack of short spacings and \textcolor{black}{the}  \textcolor{black}{lower \hi} column density detection limit. For the D--configuration the integration was relatively short (2.3 hrs) and the difference between its spectrum and AGES is primarily attributable to the difference in column density sensitivity (see below) as there should not have been any issue with detecting extended emission at scales of several arcmin as such.} 

\textcolor{black}{To gain a more complete understanding of the \hi\ in the triplet, bearing in mind the VLA flux losses relative  to AGES, we made a comparison (at common spatial and velocity resolution) between \hi\ cubes containing the triplet and the \hi\ tail from AGES and the VLA B/C-- and D--configuration. More specifically  we compared the \hi\ detected in a region within the $3\sigma$ \hi\ contour in the AGES integrated \hi\ map of FGC\,1287 and the tail, and compared this with the two VLA data cubes. This comparison revealed a substantial mass of low column density \hi\ to the east of the FGC\,1287 VLA--D \hi\ disk and in the \hi\ tail. The additional diffuse \hi\ detected in AGES in \textcolor{black}{these regions} was below the VLA--D detection limit ($\sim 2 \times 10^{19}$\,atom\,cm$^{-2}$ but above the AGES 3$\sigma$ limit of $\sim 4.5 \times 10^{17}$\,atom\,cm$^{-2}$ \citep{keenan16}. For FGC\,1287 (disk + tail) the mass of this low column density \hi\ was  estimated at $\sim$ 5 $\times$ 10$^9$ \msolar. The highest concentration of this low column density \hi\ lies to the east of the FGC\,1287  at about the same velocity ($\sim 6800$\,\km), as the mean optical velocity of the triplet (6804\,\km) 
Our analysis indicates the VLA--D, and particularly VLA--B/C configurations, are only tracing the highest column density \hi\ within a more extensive distribution of diffuse \hi\ revealed by AGES.}

\textcolor{black}{Using the VLA D--configuration we find a total \hi\ mass of the FGC\,1287 disk + tail of $9.4 \times 10^9$\,\msolar. 
Following \citep{scott10} we determined the \hi\ deficiency for FGC 1287 disk + tail by comparing this with the expected \hi\ content of similar, isolated galaxies of the same optical size and morphological type and find an \defhi\ of -0.27, i.e., an excess. If we rather use the \hi\ mass ($1.3 \times 10^{10}$\,\msolar) based on the more sensitive AGES survey the \hi\ excess rises to -0.42. Whereas an \defhi\ of -0.27 is within the range expected for gas--rich galaxies,  an \defhi\ of  -0.42 indicates either the galaxy was \textcolor{black}{particularly} \hi--rich before  producing the tail or,  more likely, a significant fraction of the \hi\  in the tail  did not originate from  FGC\,1287 but is possibly from another  triplet member and/or an \hi\ rich dwarf. Although we note the metallicity of SFC1 (see next section) is inconsistent with dwarfs as a source of its \hi,  this does not preclude them as a source of \hi\ in other parts of the tail.}

\begin{figure}
\begin{center}
\includegraphics[ angle=0,scale=0.38] {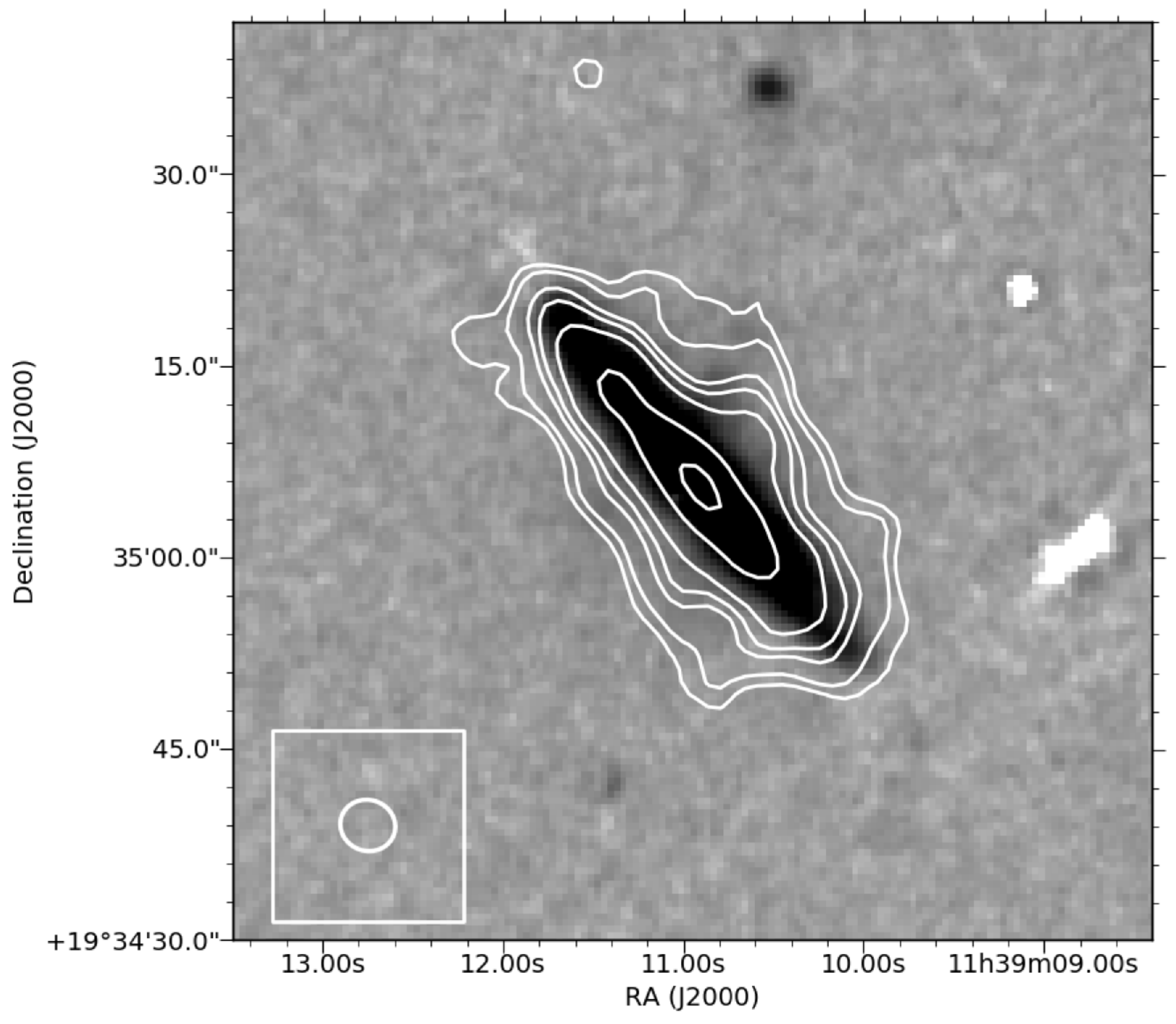}
\caption{FGC\,1287:    VLA B--configuration 1.4 GHz radio continuum contours overlaid on a smoothed  TNG \halpha+[NII] image with the lowest 3 $\sigma$ contour at 19 $\mu$Jy/beam.  The boxed ellipse at the bottom left shows the size of the  B--configuration $5.4^{\prime\prime} \times 4.9^{\prime\prime}$  VLA synthesised beam and its orientation.}
\label{fig_5}
\end{center}
\end{figure}

\begin{figure}
\begin{center}
\includegraphics[ angle=0,scale=0.32] {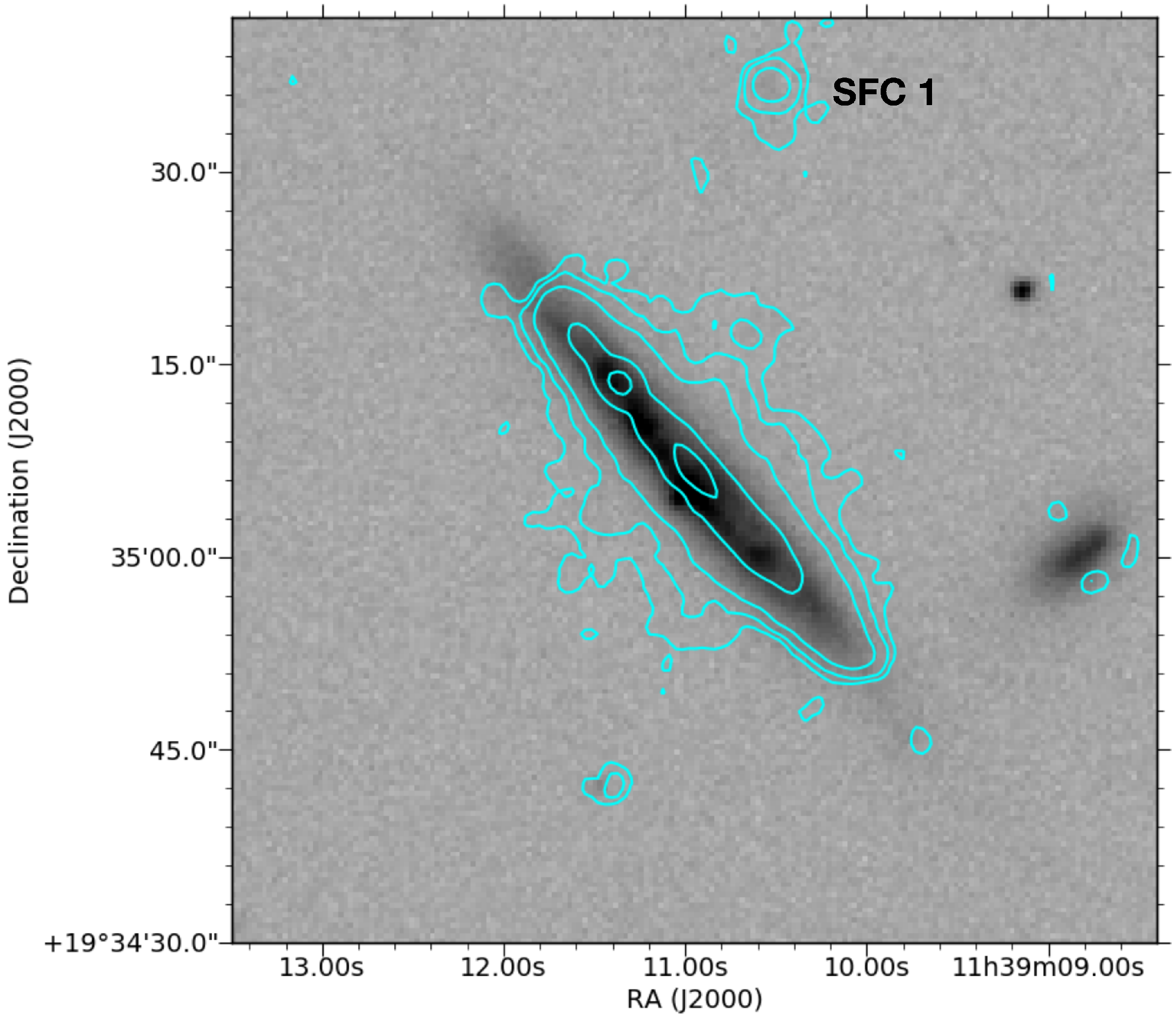}
\vspace{1cm}
\caption{FGC\,1287: TNG \halpha+[NII] contours ($3 \times 3$ boxcar smoothed) overlaid on an SDSS $i$--band image. The lowest contour is at $\sim 3\sigma$. The \textcolor{black}{location of \halpha\ clump   SFC\,1 is indicated in the figure. } }
\label{fig_6}
\end{center}
\end{figure}

Figure \ref{fig_5},  shows 1.4\, GHz radio continuum (RC) was detected in the disk of FGC\,1287 using the VLA in B--configuration (5.4 $\times$ 4.9 arcsec beam).  \textcolor{black}{The FGC\,1287  1.4 GHz RC, TNG \halpha+[NII]  (Figure \ref{fig_6})  and the nominally 145 MHz RC image \textcolor{black}{(not shown)} obtained with LOFAR (Bempong--Manful et al. in prep) all show quite similar morphologies. 
In Figure \ref{fig_5} the 1.4\, GHz RC contours display asymmetric features extending $\sim$ \textcolor{black}{10 arcsec} (4 kpc) both north and south of the FGC\,1287 disk centre,  features slightly more prominent on the north side of the object\textcolor{black}{, with LOFAR confirming this morphology.} Similar extra--planar features are seen in the \halpha+[NII]  (Figure \ref{fig_6}). We tried to look for differences in the spectral slope in the 1.4 GHz RC in the FGC\,1287 extra--planar regions which might have confirmed outflows but the signal to noise in these regions was too low to carry out such an analysis. Two further features were found to be common to \textcolor{black}{the \halpha+[NII]  and the two RC maps}. First, the  morphologies are asymmetrically distributed along the optical disk favouring the SE,   with the NE end of the each disk, where the \hi\ kinematic warp is located, being truncated relative to the optical disk (see Figure \ref{fig_6}). Second, there is no evidence of upstream compression of the emission contours that you would expect from RPS.}

We use the RC luminosity to estimate the current star formation rate (SFR) for FGC\,1287 following Equation 3  from  \cite{heesen14}, with an inclination correction, finding a value of $\sim 3.2$\,\msolaryr\ (see also Table \ref{table2}).
This is a factor of $\sim3$ higher than is observed in star--forming main sequence galaxies in the Local Volume with the same stellar mass (e.g., 0.811 \msolaryr\ \citealp{speagle14}), suggesting that the environmental process responsible for the \hi\ tail has not reduced SFR of FGC\,1287 but might have 
 recently enhanced it.


\begin{table*}
\centering
\begin{minipage}{170mm}
\caption{\hi, radio continuum  and optical properties of the VLA \hi\ detections. }
\label{table2}
\resizebox{\columnwidth}{!}{%
\begin{tabular}{@{}lrrrrrrrrrr@{}}
\hline
Galaxy ID&RA&DEC&V$_{\mathrm {HI}}$\footnote{Barycentric optical V$_{\mathrm {HI}}$ calculated using the method  described in  \cite{scott10}.}& W$_{20}$\footnote{W$_{20}$ calculated using the method described in  \cite{scott10}.} &SFR(RC)&M$_*$&Dist. from\\
&[$^h$ $^m$ $^s$] & [\degree\ \prim\ \prin\ ] &VLA--B/C&VLA--B/C&VLA-B&&FGC1287\\
& &&  [\km ] &  [\km ]&[\msolaryr]&[10$^9$\msolar]&[kpc]\\ 
\hline
FGC 1287 --disk&&&6788$\pm$3\,\,\,&335$\pm$5\,\,\,&3.2&8.59&--\\
FGC 1287--tail&&&6818$\pm$11&185$\pm$22&--&--&--\\
\hline
FGC 1287--all&11 39 10.9& +19 35 06& 6803$\pm$26&285$\pm$52&3.2 &8.59&--\\
\cg036&11 38 51.0& +19 36 05&6808$\pm$14&75$\pm$28&1.7&9.91&119\\
\cg041&11 39 24.4& +19 32 05&6786$\pm$5\,\,\,&260$\pm$10&1.5&33.26&109\\
SDSS dwarf\footnote{SDSS J113939.67+193516.7.} &11 39 39.7& +19 35 17&7381$\pm$4\,\,\,&60$\pm$7\,\,\,&--&0.002&167\\
 \hline
 \end{tabular}}
\end{minipage}
\end{table*}



\begin{table*}
\centering
\begin{minipage}{170mm}
\caption{\hi\ properties of the VLA \hi\ detections. }
\label{table2a}
\begin{tabular}{@{}lrrrrrrrrrr@{}}
\hline
Galaxy ID&M(HI)\footnote{ \mhi\ = 2.36 $\times$ 10$^2$ D$^2$ S$_{HI}$ where \mhi\ is in \msolar, D = distance (92 Mpc)  and S$_{HI}$ is the VLA B/C configuration flux  in Jy\,km\,s$^{-1}$. }  &M(HI)\footnote{From \cite{scott12}.}&MHI(HI)\footnote{Total HI flux from AGES applied using the formula in note a above. For FGC\,1287 the \hi\ flux  for the tail+disk was extracted from the AGES cube. It should be noted this AGES  \hi\ total flux  value is likely to include some side--lobe contamination from other triplet members.   }  & Def. \hi\footnote{ \hi\ deficiencies are calculated using the method as described in  \citep{scott10} and using the  \hi\ fluxes from  AGES, except for FGC\,1287 (tail+disk) which is from the VLA D--configuration \citep{scott12}.}&\af\footnote{\textcolor{black}{\af\ was measured from the \hi\ spectra as follows:  FGC\,1287 disk and SDSS dwarf from VLA B/C, \cg036 and \cg041 from AGES.} } \\
&VLA--B/C&VLA-D&AGES&&\\
& [10$^9$\msolar]&[10$^9$\msolar]&[10$^9$\msolar]&\\ 
\hline
FGC 1287 --disk&1.43&--&--&--&1.16$\pm$0.04\\
FGC 1287--tail&2.65 &--&--&--&--&\\
\hline
FGC 1287--all&4.08&9.4&13.0\,\,&-0.27&--&\\
\cg036&0.34&0.8&1.98&0.14&1.07$\pm$0.05&\\
\cg041&1.23&2.8&3.68&0.01&1.01$\pm$0.02&\\
SDSS dwarf\footnote{SDSS J113939.67+193516.7.} &0.52&--&0.67&-0.27&1.26$\pm$0.05&\\
 \hline
 \end{tabular}
\end{minipage}
\end{table*}


\subsection{Ionised gas properties}
\label{tng_result}
\textcolor{black}{As noted in Section \ref{results_hi} we  identify 5 distinct \hi\ clumps in the tail. The first three as one moves \textcolor{black}{northwards } along the \textcolor{black}{\hi} tail (SFC1, SFC2 and SFC3) coincide with blue \textcolor{black}{stellar} continuum emission in optical  imaging available as part of the DECam Legacy Survey (DECaLS), \citep{blum16}, lower left panel of Figure \ref{fig_2} (DECaLS imaging is deeper than SDSS). These three clumps are also seen as star forming  \halpha\ emitting clumps in \textcolor{black}{the} TNG narrow-band \halpha+[NII] image in the lower right panel of Figure \ref{fig_2}. Two \hi\ clumps  further \textcolor{black}{north} along the tail  (CLP4 and CLP5) are not associated with any obvious SF activity. \textcolor{black}{In Figure \ref{fig_2} -- lower left panel, the small blue galaxy just above the SW edge of  FGC\,1287 is ASK 626970 which is projected in the foreground of the triplet with an optical velocity of 1733$\pm$2  \km\ (SDSS). }  }

\begin{figure*}
\begin{center}
\includegraphics[ angle=0,scale=0.45] {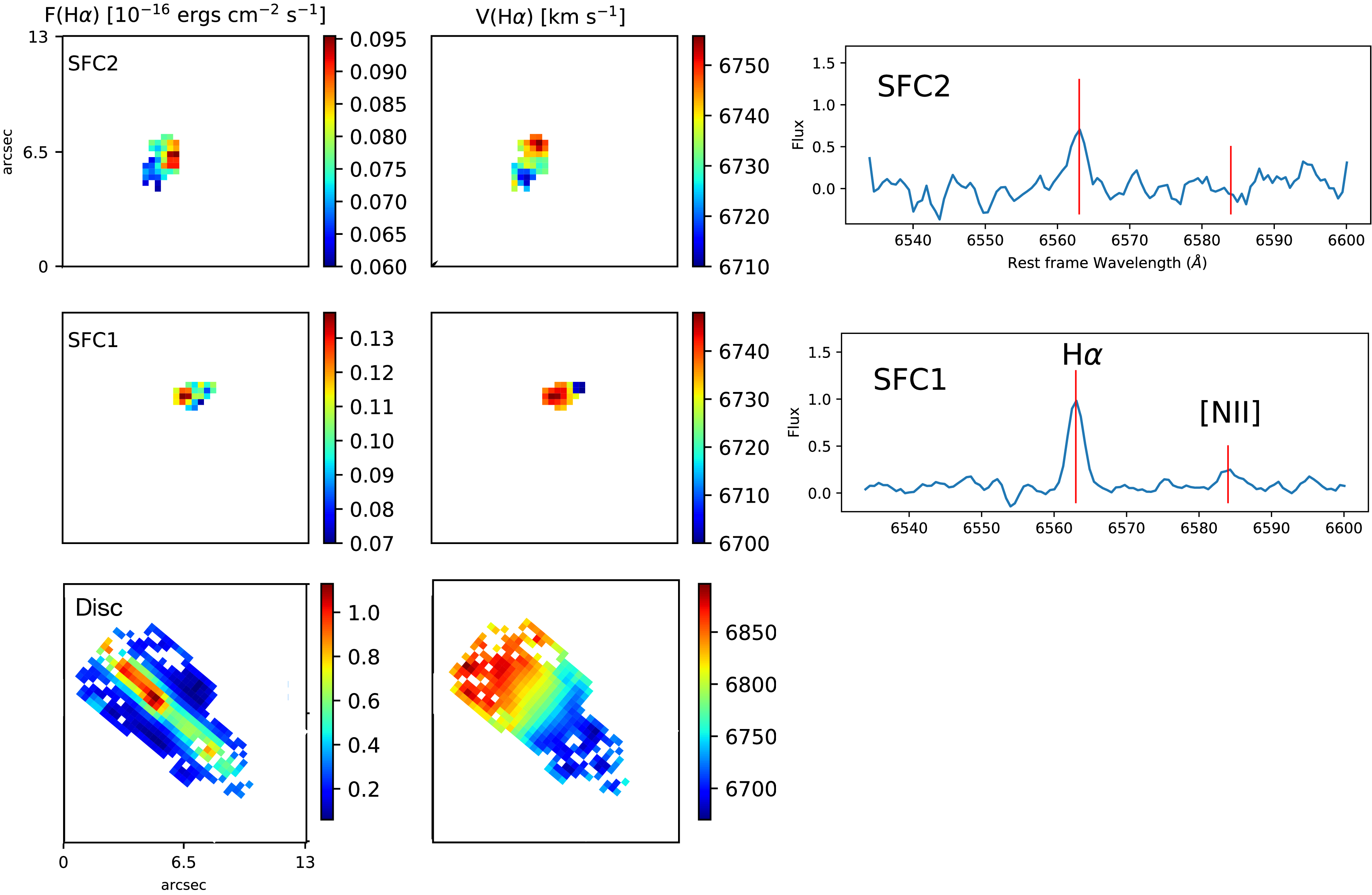}
\vspace{.01cm}
\caption{VIMOS--IFU observations of SFC1\, SFC2, and the central disk of FGC\,1287: \textit{left} Flux F(H$\alpha$) in units of 10$^{-16}$ ergs cm$^{-2}$ s$^{-1}$; \textit{middle} velocity fields V(H$\alpha$) in units of km s$^{-1}$;  \textit{right} VIMOS--IFU spectra at rest frame wavelengths. Flux is in units of 10$^{-16}$ ergs cm$^{-2}$ s$^{-1}$ \AA$^{-1}$. The red vertical lines indicate the rest frame wavelengths of the \halpha\ $\lambda$ 6563 and [NII] $\lambda$6584 emission  lines. }
\label{fig_7}
\end{center}
\end{figure*}

The star forming nature of SFC1 and SFC2 could be confirmed by our VIMOS--IFU observations as they detect clear H$\alpha$ emission coming from both clumps. The spectra for both clumps, as well as H$\alpha$ maps and velocity fields are shown in Figure \ref{fig_7}. These were obtained by fitting emission line fluxes (H$\alpha$ and [NII]$\lambda$6584) with a single Gaussian profile. The velocity field for SFC1 and SFC\,2 suggests the presence of a velocity gradient. The velocity field of the central part of FGC\,1287, basically covering the VLA B/C--configuration synthesised beam central pointing on the galaxy, shows emission from a disk in regular rotation.

The derived log([NII]$\lambda$6584/H$\alpha$) line ratios are -0.62$\pm$0.08 and -0.25$\pm$0.09 for SFC\,1 and the central FGC\,1287 disk, respectively.   
These confirm that, in both cases, ionisation comes mainly from young stars. We can thus use the Denicolo calibrator \citep{Denicolo02} 
to derive an estimate of the  oxygen abundance for both objects, finding 8.67$\pm$ 0.12 and 8.94$\pm$ 0.09 for SFC1 and FGC\,1287, respectively. This indicates  that the gas in SFC1 was pre--enriched and  most likely has its origin in the triplet and is not infalling, primordial material.

We used the relationship given in \cite{kennicut98} to estimate the star--formation rate, SFR(\halpha), for \textcolor{black}{SFC1, SFC2}  and the portion of the FGC\,1287 disk within the VIMOS field of view (FoV), after correction for galactic extinction.  No correction for internal extinction was possible as the spectrum did not cover the H$\beta$ line.
In Table \ref{table3} we show the radial velocity V$_{H\alpha}$, Flux(H$\alpha$), Flux([NII]),
12+log(O/H) and SFR(H$\alpha$) for the central FGC\,1287 disk, SFC1 and SFC2, respectively.

\begin{table*}
\footnotesize
\setlength{\tabcolsep}{4pt}
\begin{minipage}{200mm}
\caption{FGC\,1287  \halpha\ (VIMOS--IFU) and \hi\ (VLA B/C configuration) component properties}
\label{table3}
\begin{tabular}{lllcrrllrrr}
\hline
Object &\,\,\,\, RA &\,\,\,\, DEC & V$_{Ha}$ & Flux(\halpha) & Flux([NII]) &12+log & SFR(\halpha)\footnote{Corrected for  galactic extinction. } & V$_{HI}$ & W$_{20}$ & Dist.\footnote{Projected distance from the optical centre of FGC\,1287.}\\
&&&&[$\times$10$^{-16}$]&[$\times$10$^{-16}$] &(O/H)\footnote{Oxygen abundance using the \cite{Denicolo02} method.} & \\
&\,\,\,\,[$^h$ $^m$ $^s$] &\,\,\,\,[\degree\ \prim\ \prin\ ] & [\km ]&[erg\,cm$^{-2}$\,s$^{-1}$]&[erg\,cm$^{-2}$\,s$^{-1}$]&&[\msolaryr] &[\km ]&[\km ]&[kpc]\\
 \hline
disk\footnote{FGC 1287 disk only; \halpha\ values and metallicty are from  the inner part of the disk, i.e.,  from within the VIMOS--IFU $13^{\prime\prime}$ FoV. }&11 39 11.13& +19 35 07.2&6827.3&131.8$\pm$6.19&73.84$\pm3.45$&8.94$\pm$0.09&0.1105$\pm$0.0052&6788\,\,\,\,\,\,\,\,&335\,\,\,\,\,\,\,\,\,\,\,\,&-- \\
SFC 1&11 39 10.53&+19 35 36.7&6774.3&2.65$\pm$0.12&0.64$\pm$0.13&8.67$\pm$0.12&0.0022$\pm$0.0001&6728$\pm$2 &35$\pm$4\,\,\,&13\\ 
SFC 2&11 39 10.16&+19 35 55.7&6767.9&2.15$\pm$0.10& --&&0.0018$\pm$0.0001&6744$\pm$1&25$\pm$3\,\,\,&21\\
SFC 3&11 39 10.96&+19 36 52.2&--&--&--&--&--&6796$\pm$2&30$\pm$5\,\,\,&73\\
CLP 4&11 39 \,\,\,9.36&+19 39 \,\,\,7.2&--&--&--&--&--&6828$\pm$5&65$\pm$11&100\\
CLP 5&11 39 \,\,\,1.57&+19 42 32.2&--&--&--&--&--&6863$\pm$2&35$\pm$4\,\,\,&184\\
 \hline
\end{tabular}
\end{minipage}
\end{table*}

\subsection{XMM--Newton} 
\label{xmm_r}
No X--ray emission was detected in the combined 0.5 to 2\,keV band {\em XMM--Newton} image using a 5\,arcmin  radius centred on FGC\,1287. We extracted a PN spectrum and set the 2$\sigma$ limit on the normalisation of the APEC plasma model using a temperature of 0.5\,keV, with a corresponding limit on the gas mass of $7 \times 10^9$\,\msolar, assuming a uniform gas distribution. This non--detection of hot X--ray emitting IGM/ICM implies an upper limit of $n_e < 2.6 \times 10^{-5}$\,cm$^{-3}$ within a 5\,arcmin radius, which falls two orders of magnitude below the densities in the  X-ray detected ICM of Virgo, Coma, or A\,1367 \citep{bosel06b}.  Having said that, we would not have been able to detect any  warm intergalactic medium (WHIM) filaments with temperatures $\sim10^6$\,K and densities $\sim 10^{-4}$\,cm$^{-3}$ that are predicted beyond the virial radius in some cluster simulations \citep[][see also \cite{cort21}]{cen01}. \textcolor{black}{However, more typically such filaments are expected to have densities of  $\sim$ 10$^{−6}$ cm$^{−3}$}.

\textcolor{black}{We used the $\beta$ model, parameters  in  \cite{ge21} for the A\,1367 NW and SE subclusters applying them to the \cite{Cavaliere76, Gorenstein78} ICM density profile model,  \citep[][equation 10]{Reynolds21}. This gave ICM densities at the position of FGC\,1287 of $n_e$ = 1.5 $\times$ 10$^{-5}$ cm$^{-3}$  for the NW subcluster and $n_e$ = 3.1 $\times$ 10$^{-5}$ cm$^{-3}$  for the SE subcluster. In principal the sum of these densities ($n_e$ =  4.6 $\times$ 10$^{-5}$ cm$^{-3}$)  provides an estimate of the ICM density at the position of FGC 1287,  but the merging of the two subclusters means this prediction is more uncertain than for dynamically relaxed clusters. However,  the XMM ICM density upper limit is consistent to a first order with the $\beta$  model densities. } \textcolor{black}{Although, we note that using a sample of 39 galaxy clusters observed in X--ray ($ROSAT$) \cite{Vikhlinin99} found a steepening of the density profile beyond 0.3 virial radii.}
\section{DISCUSSION}
\label{discussion}

\subsection{Nature of the triplet}
\cite{scott12} showed that FGC\,1287, together with \cg41 and \cg36 are  projected \textcolor{black}{$\sim$ 1.8 Mpc west of the A\,1367 cluster centre} and between the cluster's $r_{200}$  and $r_{vir}$. \textcolor{black}{ This means they are projected within the cluster's 2.58 Mpc $r_{vir}$ \citep{moss06} and their mean optical velocity is only 313 \km\ greater than the cluster's velocity \citep{cort04}.   Given the three galaxies are projected well within the clusters virial radius and the cluster's velocity dispersion of 891 \km\ \citep{cort04}}  it is a reasonable assumption that they are bound to the cluster and have begun falling toward it, either individually or as a group. All three galaxies fall within a projected diameter of 200\,kpc, with the distance between FGC\,1287 and  each its companions $\sim 100$\,kpc or about twice the distance from us to the Magellanic Clouds, and within a  47 \km\ radial velocity range. The SDSS group catalogue by \cite{tempel12} only includes  \cg41 as a member of a pair with  FGC\,1287.

\textcolor{black}{To  first order the optical DECaLS DR 5 images for FGC\,1287, \cg036 and \cg041 present symmetric morphologies which rules out  recent ($\lesssim$1 Gyr) strong tidal interactions. However, all three galaxies display second order asymmetries which could be either intrinsic characteristics or signatures of earlier major or recent minor interactions, e.g., FGC\,1287’s high--intensity $g$--band morphology is more variable in the NW of the disk compared to the SW consistent with the disturbance suggested by the \hi\ velocity field. } 

The transverse velocities of the three galaxies are not directly constrained by our observations. However,  \textcolor{black}{assuming the tail originated from FGC\,1287, the geometry of the \hi\ tail suggests a  transverse motion for FGC\,1287 in the N--S direction,  with a relatively low velocity in the E--W direction. Although the  \hi\ tail appears in the VLA moment 0 maps to belong to FGC\,1287, we consider two possible origins for the tail. }  If the  long \hi\ tail  consists of  \hi\  stripped directly from the FGC\,1287 disk by RPS, referred to from here on the direct stripping scenario,  it implies FGC\,1287  is an interloper relative to the other two galaxies with a large  N--S transverse velocity. On the other hand, the tail could consist of triplet \textcolor{black}{cold} IGM previously tidally stripped during interactions between its members and recently  removed by RPS and this is referred to from here on as the \textcolor{black}{combined gravitational and hydrodynamic}  scenario.  \textcolor{black}{Under either scenario the motion generating the tail appears to be principally perpendicular to radius vector pointing toward the cluster centre (see arrow in Figures \ref{fig_2}, top left panel and \ref{fig_2a})}. 

The  12+log(O/H)  oxygen abundance from our VIMOS--IFU analysis for SFC\,1 is 8.67$\pm$0.12 and 8.94$\pm$0.09 for the central 27 arcsec$^2$ of the FGC\,1287 disk. Additionally, we derived    12+log(O/H) for the inner 3 arcsec diameter of the disk from SDSS spectra  for FGC\, 1287, \cg036  and  \cg041  of  8.81 $\pm$ 0.08,  9.06 and 8.79 respectively.  
Whereas the VIMOS 12+log(O/H) abundance for the FGC\,1287 central disk is, within the errors,  higher than in SFC\,1,  the uncertainties due to the limited disk areas sampled mean we are not able to distinguish SFC1's metallicity from that of any of the group members, i.e., the presence of pre--enriched material in SFC1  is expected under either RPS scenario. In the following two subsections we consider the evidence for and against the each of these RPS scenarios. 
 
\subsection{FGC\,1287 direct RPS scenario}
 The long, narrow \hi\ tail emanating from FGC\,1287 \textcolor{black}{in the VLA moment 0 maps}   is consistent with a morphology predicted in simulations for strong ram pressure stripping, \textcolor{black}{ICM $n_e$  }=  $\sim$  10$^{-3}$ cm$^{-3}$ and \vrel\ 1000 \km\ near a cluster centre \citep{lee20}. Given the {\em XMM--Newton} results indicate a significantly lower ICM density ($n_e < 2.6 \times 10^{-5}$\,cm$^{-3}$), it follows if RPS produced the FGC\,1287 \hi\ tail,  it must have involved either an unusually high \vrel\ or an extremely weak  gravitational restoring force binding the \hi\  to the galaxy compared with the parameters in the simulations by \cite{lee20}.  The dynamical mass (M$_{dym}$) within the \hi\ disk radius is $\sim$ 8.8 $\times$ 10$^{10}$ \msolar\ and being within the canonical range for a late--type spiral galaxy, does not suggest abnormally  weak gravitational restoring forces in the FGC\,1287 disk. 

RPS models \citep[e.g.,][]{roed07} predict outside--in truncation of \hi\ disks. Under the direct stripping scenario the \hi\ now in FGC\,1287's tail was orginally part of its \hi\ disk. The VLA--D configuration disk+tail \mhi\ $= 9.4 \times 10^9$\,\msolar\ \citep{scott12} \textcolor{black}{which} provides an estimate of the FGC\,1287 disk's \mhi\ pior to RPS. Using this \hi\ mass, the expected \hi\ disk diameter, D$_{HI}$, = 56$\pm$1\,kpc based on the D$_{HI}$/M$_{HI}$ relation from \cite{wang16}.  FGC\,1287's  current \hi\ disk diameter\footnote{at the $1.25 \times 10^{20}$\,atoms\,cm$^{-2}$ contour level which is equivalent to $\Sigma_{HI}$ = 1\,\msolar\,pc$^{-2}$.}   is 1.02$\pm$0.23 arcmin (25.3$\pm$6\,kpc)  indicating that its  \hi\ disk is significantly truncated. Confirmation of the truncation of the FGC\,1287 \hi\ disk is seen in Figure \ref{fig_3} where the lowest \hi\ column density contour approximately coincides with the optical disk edges. 

\begin{figure}
\begin{center}
\includegraphics[ angle=0,scale=0.56] {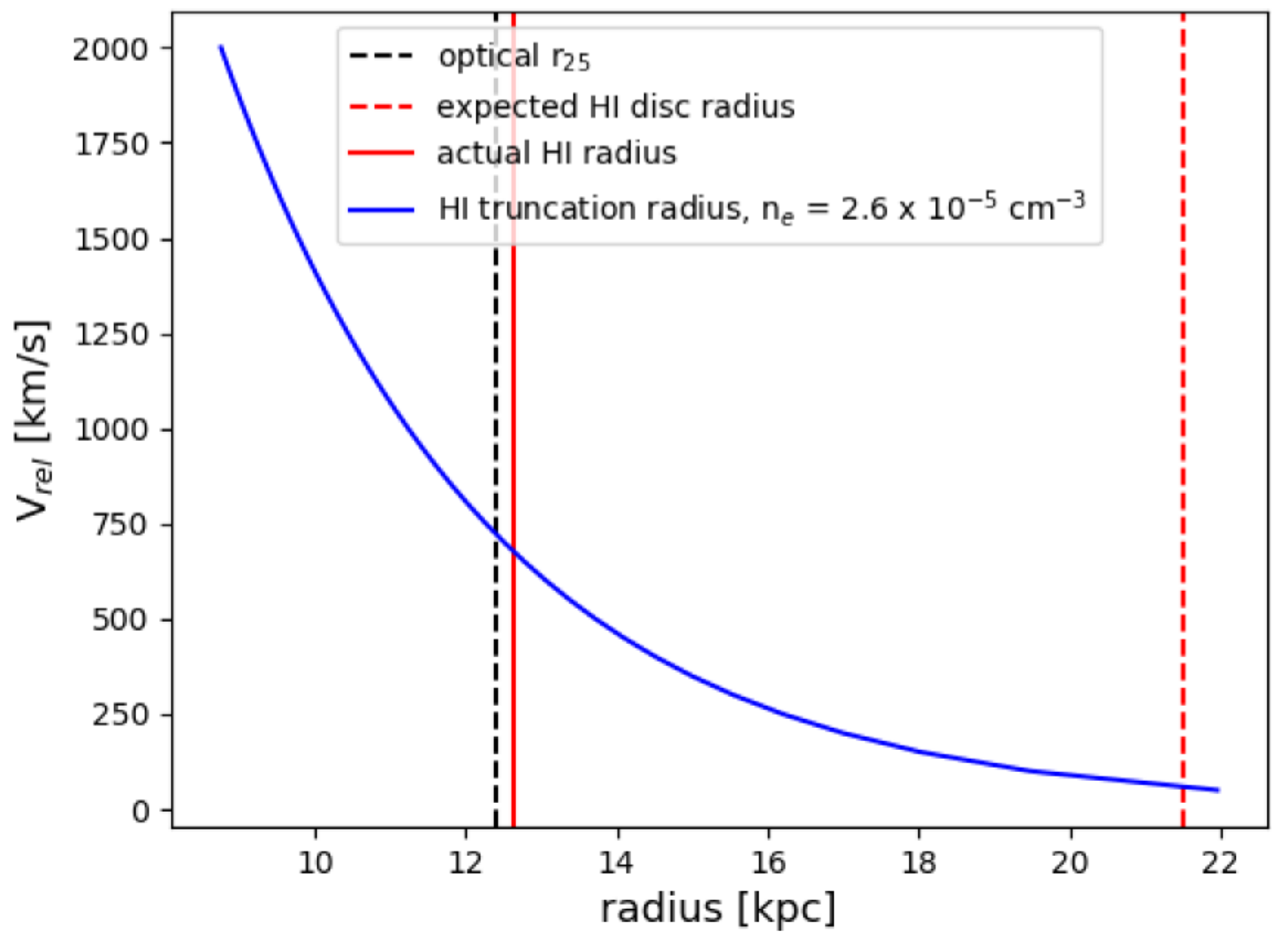}
\vspace{1cm}
\caption{FGC\,1287 \hi\  truncation radius from Steinhauser et al. (2016) in kpc as a function of  \vrel\ for an ICM number density upper limit from the {\em XMM--Newton} X--ray observation of $ n_e = 2.6 \times 10^{-5}$ \textcolor{black}{cm$^{-3}$} (blue curve). The vertical lines identified in the legend indicate that the \hi\ disk has been truncate down to the optical $r_{25}$radius. }
\label{fig8}
\end{center}
\end{figure}

An alternative way to evaluate the direct stripping scenario of the FGC\,1287 \hi\ disk is to look at the combination of velocity and \textcolor{black}{ ICM $n_e$} required to truncate the FGC\,1287 \hi\ disk to its observed radius, given the upper limit for  ICM $n_e$ from {\em XMM--Newton}. Figure \ref{fig8} shows as vertical lines the optical $r_{25}$, the observed \hi\ radius and the expected \hi\ radius, quantified above. The plot's blue curve shows the \hi\ stripping radius using Equation 8 from \cite{Steinhauser16}  for \vrel\ with the upper limit from {\em XMM--Newton} \textcolor{black}{for ICM $n_e$}  = 2.6 $\times$  10$^{-5}$ \textcolor{black}{cm$^{-3}$}. The  plot  shows a \vrel\ of $\sim 700$\,\km\ is required to explain the \hi\ disk truncation at the {\em XMM--Newton} \textcolor{black}{ ICM $n_e$} upper limit in terms of the direct stripping scenario. \textcolor{black}{However the substantial mass of low column density \hi, below the VLA--D detection limit and to the East of FGC\,1287's disk is inconsistent with the truncated disk and narrow tail morphology expected from RPS models, at least with a \vrel\ of this magnitude.}
 
 A major difficulty for this scenario is explaining how FGC\,1287 acquired a \vrel\ of $\gtrsim$ 700 \km, in such a low galaxy density environment. This \vrel\ is a factor of $\gtrsim$3 higher than expected in groups and filaments.   Moreover, if this \vrel\ ICM $n_e$ combination is capable of displacing $>$50\%  of an LTGs \hi\ mass,  it is hard to understand why  LTGs closer to the cluster core still contain significant \hi\ masses This is because RPS is expected to be much more effective closer to the cluster due to both higher \textcolor{black}{ ICM $n_e$}  and \vrel\ consistent with the cluster's velocity dispersion of 891 \km.
 
 \textcolor{black}{Another possible explanation for the \hi\ tail is that it was caused by a cluster merger shock compressing the ICM downstream of the tail. There is no indication of synchrotron emission from such a shock within a 30 arcmin radius ($\sim$ 745 kpc at the distance of A\,1367) of FGC\,1287 in the LOFAR image. Moreover, the orientation of the \hi\ tail is perpendicular, rather than parallel, to the direction in which such a shock would be expected to propagate, i.e., outward from the cluster centre.}

\subsection{Combined gravitational and hydrodynamic stripping scenario}
Under the \textcolor{black}{combined gravitational and hydrodynamic stripping scenario}  the weakly bound \textcolor{black}{cold} IGM\textcolor{black}{, principally \hi,} accumulated from multiple past intra--group interactions has recently been subject to RPS \textcolor{black}{as the group fell towards A\,1367 and came}  into contact with  that cluster's warm ICM.  This scenario requires a much lower \vrel, at the same  ICM $n_e$, to remove \hi\ compared to the  direct stripping scenario. \textcolor{black}{This is because the IGM is only loosely bound to the gravitation potential of the group compared to the more tightly bound ISM in galaxy  disks.} The presence of a large mass of low column density \hi, below the VLA detection limit,  at velocities close to the mean optical velocity of the triplet  and near the projected centre of mass of the triplet  is consistent with this scenario. In this scenario  the \hi\ tail would be analogous to the 330\,kpc  ionised \halpha\ tail of presumably IGM material emanating from the Blue Infalling Group (BIG) in A\,1367 \citep{yagi17}. For this scenario it is assumed that the triplet, possibly including SDSS J113939.67+193516.7,  is a bound group. 

FGC\,1287 is a highly inclined galaxy and support for the \textcolor{black}{combined gravitational and hydrodynamic stripping scenario} comes from \textcolor{black}{the  unusual AGES and} VLA--D configuration \hi\ profiles for  the  FGC\,1287 disk+tail in Figure \ref{fig_4}. \hi--rich high inclination LTGs typically present double horn \hi\ profiles because of \textcolor{black}{their rotating  \hi\ disks}. Two  high inclination A\,1367 LTGs  (\cg087 and \cg121) projected within 550 kpc of the cluster centre and undergoing RPS both display one--sided tails in resolved \hi\ maps.  Figure \ref{figb} in Appendix B, shows the AGES \hi\ profiles for these high inclination galaxies  displaying double horned \hi\ profiles, with the principal \hi\ profile asymmetry consisting of one horn being significantly smaller than the other, i.e., ram pressure is preferentially impacting the outer \hi\ disks. Although the AGES FGC\,1287 disk+tail  spectrum does have  secondary maxima which  align with the double horns of the FGC\,1287 \hi\ disk, see Figure \ref{fig_4}, these are dwarfed by the central maximum which is close to the FGC\,1287 \hi\ systemic velocity and the optical velocities of the \cg036 and \cg041 companions. Moreover,  the D--configuration  \hi\ maxima align in velocity with the velocities of the \hi\ clumps in the tail, rather than the velocities at the edges of the remaining truncated FGC\,1287 \hi\ disk, as would be expected from direct RPS.        So, past interactions between triplet members followed by IGM RPS stripping could explain the AGES and VLA D--configuration profiles and tail clump velocities. 

\subsection{Filament transit}
FGC\,1287 is projected   $\sim$ 30 arcmin (740 kpc) south of a filament of galaxies joining the cluster from the NW (see Figure 8 in \cite{scott18}).  So, ram pressure stripping during a N to S passage of a gas filament associated with this galaxy filament is a possible mechanism for generating the \hi\ tail. To produce an \hi\ tail as impressive as that seen in FGC\,1287 by direct ram \textcolor{black}{pressure} stripping we would expect a ram pressure force comparable to that of most extreme ongoing ram pressure stripping cases near cluster cores, i.e., ram pressure from a combination of  gas densities comparable to ICM  in  rich clusters cores ($n_e$ = 3 $\times$ 10$^{-3}$ cm$^3$) at a \vrel\  of at least a typical cluster velocity dispersion $\sim$ 1000 \km). Even at a \vrel\  of 750 \km (which would be hard to explain in a filament where the expected velocity dispersion is $\sim$ 250 \km), the time required to cover the projected distance between FGC\,1287 and the filament would be $\sim$ 1 Gyr during which time the \hi\ in the tail would be expected to have either been lost or fallen back into the group or FGC\,1287 gravitational potential. As this is  not what is observed,  this mechanism seems unlikely.

\subsection{Evidence for a recent interaction}
 FGC\,1287's warped \hi\ disk (Figure \ref{fig_3}) presents the clearest evidence of an interaction within the last $\sim$ 0.7\,Gyr, the \hi\ relaxation time--scale \textcolor{black}{for major mergers} \citep{holwerda11}. The closed velocity contours in the velocity field of \cg041, at velocities similar to the warp in FGC\,1287, and its morphology, being more extended towards the East (Figure \ref{fig_3}) could be the result of an interaction with FGC\,1287. The time scale for this interaction assuming a separation velocity of $\sim$250 \km\ would be $\sim$ 0.4\,Gyr, i.e., within the \hi\ relaxation time scale for both galaxies. However, this does not provide definitive evidence of a recent tidal interaction.

Supporting evidence for a recent interaction comes from the star formation history derived from the FGC\,1287 SDSS spectrum using the Fitting Analysis using Differential evolution Optimization (FADO) spectral synthesis code \citep{gomes17}, which shows a spike of star formation a few $\times$ 10$^7$ years ago.  Unfortunately, this interaction time scale is not precise enough to discriminate between the direct and combined gravitational and hydrodynamic stripping scenarios. For now, all we can say is that the FGC\,1287 \hi\ disk has suffered a recent and strong interaction which could have been either tidal or hydrodynamic.

\textcolor{black}{There are examples of galaxy flybys creating long HI tails in the literature, e.g, \cite{serra13}. Although, in that case, the \hi\ tail mass is an order of magnitude lower and more diffuse (without star formation) than for FGC\,1287. In the \cite{serra13}  case, tidal \hi\ stripping was thought to arise from an encounter between NGC\,3187 with either the gravitational potential of the group or the group member NGC\,3162. For the FGC\,1287 \hi\ tail we could not identify a flyby companion. However, an interaction amongst the triplet members (possibly one of them joining the other two) with the gravitational potential of the group might explain the observed kinematics and the change in projected direction midway along the tail.}

\subsection{Star formation in the \hi\ tail}

Perhaps one of the most unexpected results of our analysis is that the VIMOS observations confirmed SF within two of the \hi\ tail clumps, SFC1 and SFC2. Both tail clumps have \hi\ column densities $>$ 3.1 $\times$ 10$^{20}$ atoms cm$^{-2}$, i.e., above the threshold for SF reported in \cite{maybhate07}. The stars in the FGC\,1287 \hi\ tail are almost certainly formed via in--situ gravitational gas collapse. Given the large \hi\ mass in the FGC\,1287 tail, the SFE in the tail appears to be orders of magnitude lower than observed in cluster core Jellyfish/RPS galaxies, e.g., A2667 and A1689, where the stellar tails are prominent in optical images. The presence of SF is an interesting aspect of the long \hi\ tail, but it is important to note this is happening away from the relatively highly pressurised ICM  cluster core environments where SF in  Jellyfish/RPS  tails are normally observed. Near cluster cores the high ICM pressure appears to promote the conversion of \hi\  in the tails to the molecular phase which in turn promotes star formation in the tails \citep{jachym17,jachym19,Chen20}. \textcolor{black}{Although,  the exceptional $\sim$ 90 kpc multi--phase RPS tail of JO206 in the low masss cluster IIZW 108 suggested that high \vrel\ and the resulting draping of the ICM magnetic field may be a more important factor than ICM density in promoting star formation in RPS tails \citep{muller21}. ICM draping is the process by which as a LTG penetrates a clusters ICM the magnetic fields in the ICM  envelope and align around the LTG and it's RPS tail.}

\section{Concluding remarks}
\label{concl}
In summary, our new observations reveal FGC\,1287 has undergone an interaction within the last 0.7 Gyr, but the type of interaction, tidal or RPS, remains unclear.   Our \halpha+[NII]  imaging shows a chain of at least three star--forming knots along the base of the \hi\ tail.  FGC\,1287 presents the primary signatures of RPS, i.e.,  a one--sided ionised or neutral gas tail and unperturbed stellar disk.  However, the key prerequisite for RPS is the presence of ICM with sufficient density to strip \hi\  from a galaxy's disk at the galaxy's \vrel.  One--sided RPS tails are normally observed projected in proximity to an ICM with sufficient density to be detected at X--ray wavelengths.   But in the case of FGC\,1287  it is projected more than one megaparsec from the nearest X--ray detected ICM and the non--detection of X--ray emission from our {\em XMM--Newton} pointing at the triplet position implies a \textcolor{black}{ ICM $n_e$} $<$ 2.6 $\sim$ 10$^{-5}$ \textcolor{black}{cm$^{-3}$}.  In a scenario in which the \hi\ in the FGC\,1287 disk is directly removed by RPS the {\em XMM--Newton} upper limit to \textcolor{black}{ ICM $n_e$} implies a \vrel\  $>$ 700 \km. Moreover, although a recent onset of RPS could be explained as the result of the galaxy's recent arrival at the cluster outskirts it hard to envisage how this could have been sufficiently strong to remove the large mass of \hi\ now observed in the tail. 

Then, as discussed in \cite{scott12}, the geometry of the FGC\,1287 tail and the lack of signs of strong recent disturbances observed in the optical disks of the other two galaxies make a purely tidal origin for the FGC\,1287 tail unlikely as well. A hybrid explanation of pre-processing within the triplet followed by ram pressure stripping of the triplet's poorly bound IGM to produce the \hi\ tail, while speculative, seems a plausible alternative.
 
This object remains an enigma and is a reminder of how complex the task of isolating environmental processes affecting galaxy evolution is, even for galaxies in relatively moderately dense environments such as filaments and small groups.

\section*{Acknowledgements}
\textcolor{black}{We would like to thank the referee for several useful comments which improved the clarity of the manuscript.} TS and PL  acknowledge support by Funda\c{c}\~{a}o para a Ci\^{e}ncia e a Tecnologia (FCT) through national funds (UID/FIS/04434/2013), FCT/MCTES through national funds (PIDDAC) by this grant UID/FIS/04434/2019 and by FEDER through COMPETE2020 (POCI--01--0145--FEDER--007672).  TS and PL also acknowledges support from DL 57/2016/CP1364/CT0009 and DL 57/2016/CP1364/CT0010. \textcolor{black}{LC acknowledges support from the Australian Research Council’s Discovery Project and Future Fellowship funding schemes (DP210100337,FT180100066). Parts of this research were conducted by the Australian Research Council Centre of Excellence for All Sky Astrophysics in 3 Dimensions (ASTRO 3D), through project number CE170100013.} This research has made use of the NASA/IPAC Extragalactic Database (NED) which is operated by the Jet Propulsion Laboratory, California Institute of Technology, under contract with the National Aeronautics and Space Administration.This research has made use of the Sloan Digital Sky Survey (SDSS). The SDSS Web Site is http://www.sdss.org/.
IRAF is distributed by the National Optical Astronomy Observatory, which is operated by the Association of Universities for Research in Astronomy (AURA) under a cooperative agreement with the National Science Foundation. This research made use of APLpy, an open--source plotting package for Python \citep{Robitaille2012}.

\section*{DATA AVAILABILITY}

The data underlying this article will be shared on reasonable request to the corresponding author.

\bibliographystyle{mnras}
\bibliography{cluster}

\begin{thebibliography}{}
\makeatletter
\relax
\def\mn@urlcharsother{\let\do\@makeother \do\$\do\&\do\#\do\^\do\_\do\%\do\~}
\def\mn@doi{\begingroup\mn@urlcharsother \@ifnextchar [ {\mn@doi@}
  {\mn@doi@[]}}
\def\mn@doi@[#1]#2{\def\@tempa{#1}\ifx\@tempa\@empty \href
  {http://dx.doi.org/#2} {doi:#2}\else \href {http://dx.doi.org/#2} {#1}\fi
  \endgroup}
\def\mn@eprint#1#2{\mn@eprint@#1:#2::\@nil}
\def\mn@eprint@arXiv#1{\href {http://arxiv.org/abs/#1} {{\tt arXiv:#1}}}
\def\mn@eprint@dblp#1{\href {http://dblp.uni-trier.de/rec/bibtex/#1.xml}
  {dblp:#1}}
\def\mn@eprint@#1:#2:#3:#4\@nil{\def\@tempa {#1}\def\@tempb {#2}\def\@tempc
  {#3}\ifx \@tempc \@empty \let \@tempc \@tempb \let \@tempb \@tempa \fi \ifx
  \@tempb \@empty \def\@tempb {arXiv}\fi \@ifundefined
  {mn@eprint@\@tempb}{\@tempb:\@tempc}{\expandafter \expandafter \csname
  mn@eprint@\@tempb\endcsname \expandafter{\@tempc}}}

\bibitem[\protect\citeauthoryear{{Bahcall}}{{Bahcall}}{1999}]{bacall99}
{Bahcall} N.~A.,  1999, in {Dekel} A.,  {Ostriker} J.~P.,  eds, Formation of
  Structure in the Universe. p.~135

\bibitem[\protect\citeauthoryear{{Bellhouse} et~al.,}{{Bellhouse}
  et~al.}{2019}]{bellhouse19}
{Bellhouse} C.,  et~al., 2019, \mn@doi [MNRAS] {10.1093/mnras/stz460}, \href
  {https://ui.adsabs.harvard.edu/abs/2019MNRAS.485.1157B} {485, 1157}

\bibitem[\protect\citeauthoryear{{Blum} et~al.,}{{Blum} et~al.}{2016}]{blum16}
{Blum} R.~D.,  et~al., 2016, in American Astronomical Society Meeting Abstracts
  \#228. p. 317.01

\bibitem[\protect\citeauthoryear{{Boselli} \& {Gavazzi}}{{Boselli} \&
  {Gavazzi}}{2006}]{bosel06a}
{Boselli} A.,  {Gavazzi} G.,  2006, \mn@doi [PASP] {10.1086/500691}, \href
  {http://adsabs.harvard.edu/abs/2006PASP..118..517B} {118, 517 }

\bibitem[\protect\citeauthoryear{{Boselli}, {Boissier}, {Cortese}, {Gil de
  Paz}, {Seibert}, {Madore}, {Buat}  \& {Martin}}{{Boselli}
  et~al.}{2006}]{bosel06b}
{Boselli} A.,  {Boissier} S.,  {Cortese} L.,  {Gil de Paz} A.,  {Seibert} M.,
  {Madore} B.~F.,  {Buat} V.,   {Martin} D.~C.,  2006, \mn@doi [ApJ]
  {10.1086/507766}, \href {http://adsabs.harvard.edu/abs/2006ApJ...651..811B}
  {651, 811}

\bibitem[\protect\citeauthoryear{{Boselli} et~al.,}{{Boselli}
  et~al.}{2016}]{bosel16}
{Boselli} A.,  et~al., 2016, \mn@doi [A\&A] {10.1051/0004-6361/201527795},
  \href {http://esoads.eso.org/abs/2016A%26A...587A..68B} {587, A68}

\bibitem[\protect\citeauthoryear{{Bravo-Alfaro}, {Cayatte}, {van Gorkom}  \&
  {Balkowski}}{{Bravo-Alfaro} et~al.}{2000}]{bravo00}
{Bravo-Alfaro} H.,  {Cayatte} V.,  {van Gorkom} J.~H.,   {Balkowski} C.,  2000,
  \mn@doi [AJ] {10.1086/301194}, \href
  {http://adsabs.harvard.edu/abs/2000AJ....119..580B} {119, 580}

\bibitem[\protect\citeauthoryear{{Cavaliere} \& {Fusco-Femiano}}{{Cavaliere} \&
  {Fusco-Femiano}}{1976}]{Cavaliere76}
{Cavaliere} A.,  {Fusco-Femiano} R.,  1976, A\&A, \href
  {https://ui.adsabs.harvard.edu/abs/1976A&A....49..137C} {500, 95}

\bibitem[\protect\citeauthoryear{{Cen}, {Tripp}, {Ostriker}  \&
  {Jenkins}}{{Cen} et~al.}{2001}]{cen01}
{Cen} R.,  {Tripp} T.~M.,  {Ostriker} J.~P.,   {Jenkins} E.~B.,  2001, \mn@doi
  [ApJL] {10.1086/323721}, \href
  {http://esoads.eso.org/abs/2001ApJ...559L...5C} {559, L5}

\bibitem[\protect\citeauthoryear{{Chen} et~al.,}{{Chen} et~al.}{2020}]{Chen20}
{Chen} H.,  et~al., 2020, \mn@doi [MNRAS] {10.1093/mnras/staa1868}, \href
  {https://ui.adsabs.harvard.edu/abs/2020MNRAS.496.4654C} {496, 4654}

\bibitem[\protect\citeauthoryear{{Chung}, {van Gorkom}, {Kenney}  \&
  {Vollmer}}{{Chung} et~al.}{2007}]{chung07}
{Chung} A.,  {van Gorkom} J.~H.,  {Kenney} J.~D.~P.,   {Vollmer} B.,  2007,
  \mn@doi [ApJL] {10.1086/518034}, \href
  {http://adsabs.harvard.edu/abs/2007ApJ...659L.115C} {659, L115}

\bibitem[\protect\citeauthoryear{{Chung}, {van Gorkom}, {Kenney}, {Crowl}  \&
  {Vollmer}}{{Chung} et~al.}{2009}]{chung09}
{Chung} A.,  {van Gorkom} J.~H.,  {Kenney} J.~D.~P.,  {Crowl} H.,   {Vollmer}
  B.,  2009, \mn@doi [AJ] {10.1088/0004-6256/138/6/1741}, \href
  {http://adsabs.harvard.edu/abs/2009AJ....138.1741C} {138, 1741}

\bibitem[\protect\citeauthoryear{{Consolandi}, {Gavazzi}, {Fossati},
  {Fumagalli}, {Boselli}, {Yagi}  \& {Yoshida}}{{Consolandi}
  et~al.}{2017}]{cosolandi17}
{Consolandi} G.,  {Gavazzi} G.,  {Fossati} M.,  {Fumagalli} M.,  {Boselli} A.,
  {Yagi} M.,   {Yoshida} M.,  2017, \mn@doi [\aap]
  {10.1051/0004-6361/201731218}, \href
  {http://esoads.eso.org/abs/2017A%26A...606A..83C} {606, A83}

\bibitem[\protect\citeauthoryear{{Cortese}, {Gavazzi}, {Boselli},
  {Iglesias-Paramo}  \& {Carrasco}}{{Cortese} et~al.}{2004}]{cort04}
{Cortese} L.,  {Gavazzi} G.,  {Boselli} A.,  {Iglesias-Paramo} J.,   {Carrasco}
  L.,  2004, \mn@doi [A\&A] {10.1051/0004-6361:20040381}, \href
  {http://adsabs.harvard.edu/abs/2004A%26A...425..429C} {425, 429 }

\bibitem[\protect\citeauthoryear{{Cortese}, {Gavazzi}, {Boselli}, {Franzetti},
  {Kennicutt}, {O'Neil}  \& {Sakai}}{{Cortese} et~al.}{2006}]{cort06}
{Cortese} L.,  {Gavazzi} G.,  {Boselli} A.,  {Franzetti} P.,  {Kennicutt}
  R.~C.,  {O'Neil} K.,   {Sakai} S.,  2006, \mn@doi [A\&A]
  {10.1051/0004-6361:20064873}, \href
  {http://adsabs.harvard.edu/abs/2006A%26A...453..847C} {453, 847}

\bibitem[\protect\citeauthoryear{{Cortese} et~al.,}{{Cortese}
  et~al.}{2007}]{cort07}
{Cortese} L.,  et~al., 2007, \mn@doi [MNRAS]
  {10.1111/j.1365-2966.2006.11369.x}, \href
  {http://adsabs.harvard.edu/abs/2007MNRAS.376..157C} {376, 157}

\bibitem[\protect\citeauthoryear{{Cortese} et~al.,}{{Cortese}
  et~al.}{2008}]{cort08}
{Cortese} L.,  et~al., 2008, \mn@doi [MNRAS]
  {10.1111/j.1365-2966.2007.12664.x}, \href
  {http://adsabs.harvard.edu/abs/2008MNRAS.383.1519C} {383, 1519}

\bibitem[\protect\citeauthoryear{{Cortese} et~al.,}{{Cortese}
  et~al.}{2012}]{cort12}
{Cortese} L.,  et~al., 2012, \mn@doi [A\&A] {10.1051/0004-6361/201219312},
  \href {http://adsabs.harvard.edu/abs/2012A%26A...544A.101C} {544, A101}

\bibitem[\protect\citeauthoryear{{Cortese}, {Catinella}  \& {Smith}}{{Cortese}
  et~al.}{2021}]{cort21}
{Cortese} L.,  {Catinella} B.,   {Smith} R.,  2021, \mn@doi [PASA]
  {10.1017/pasa.2021.18}, \href
  {https://ui.adsabs.harvard.edu/abs/2021PASA...38...35C} {38, e035}

\bibitem[\protect\citeauthoryear{{Denicol{\'o}}, {Terlevich}  \&
  {Terlevich}}{{Denicol{\'o}} et~al.}{2002}]{Denicolo02}
{Denicol{\'o}} G.,  {Terlevich} R.,   {Terlevich} E.,  2002, \mn@doi [MNRAS]
  {10.1046/j.1365-8711.2002.05041.x}, \href
  {http://cdsads.u-strasbg.fr/abs/2002MNRAS.330...69D} {330, 69}

\bibitem[\protect\citeauthoryear{{Di Teodoro} \& {Fraternali}}{{Di Teodoro} \&
  {Fraternali}}{2015}]{DiTeodoro15}
{Di Teodoro} E.~M.,  {Fraternali} F.,  2015, \mn@doi [MNRAS]
  {10.1093/mnras/stv1213}, \href
  {http://adsabs.harvard.edu/abs/2015MNRAS.451.3021D} {451, 3021}

\bibitem[\protect\citeauthoryear{{Dressler}}{{Dressler}}{2004}]{dress04}
{Dressler} A.,  2004, in {Mulchaey} J.~S.,  {Dressler} A.,   {Oemler} A.,  eds,
  Clusters of Galaxies: Probes of Cosmological Structure and Galaxy Evolution.
  p.~206

\bibitem[\protect\citeauthoryear{{Gavazzi} \& {Jaffe}}{{Gavazzi} \&
  {Jaffe}}{1987}]{gava87}
{Gavazzi} G.,  {Jaffe} W.,  1987, A\&A, \href
  {http://adsabs.harvard.edu/abs/1987A%26A...186L...1G} {186, L1}

\bibitem[\protect\citeauthoryear{{Gavazzi}, {Contursi}, {Carrasco}, {Boselli},
  {Kennicutt}, {Scodeggio}  \& {Jaffe}}{{Gavazzi} et~al.}{1995}]{gava95}
{Gavazzi} G.,  {Contursi} A.,  {Carrasco} L.,  {Boselli} A.,  {Kennicutt} R.,
  {Scodeggio} M.,   {Jaffe} W.,  1995, A\&A, \href
  {http://adsabs.harvard.edu/abs/1995A%26A...304..325G} {304, 325}

\bibitem[\protect\citeauthoryear{{Gavazzi}, {Boselli}, {Mayer},
  {Iglesias-Paramo}, {V{\'{\i}}lchez}  \& {Carrasco}}{{Gavazzi}
  et~al.}{2001}]{gava01b}
{Gavazzi} G.,  {Boselli} A.,  {Mayer} L.,  {Iglesias-Paramo} J.,
  {V{\'{\i}}lchez} J.~M.,   {Carrasco} L.,  2001, \mn@doi [ApJL]
  {10.1086/338389}, \href {http://adsabs.harvard.edu/abs/2001ApJ...563L..23G}
  {563, L23}

\bibitem[\protect\citeauthoryear{{Ge} et~al.,}{{Ge} et~al.}{2021}]{ge21}
{Ge} C.,  et~al., 2021, \mn@doi [MNRAS] {10.1093/mnras/stab1569}, \href
  {https://ui.adsabs.harvard.edu/abs/2021MNRAS.505.4702G} {505, 4702}

\bibitem[\protect\citeauthoryear{{Gomes} \& {Papaderos}}{{Gomes} \&
  {Papaderos}}{2017}]{gomes17}
{Gomes} J.~M.,  {Papaderos} P.,  2017, \mn@doi [A\&A]
  {10.1051/0004-6361/201628986}, \href
  {http://cdsads.u-strasbg.fr/abs/2017A%26A...603A..63G} {603, A63}

\bibitem[\protect\citeauthoryear{{Gorenstein}, {Fabricant}, {Topka}, {Harnden}
  \& {Tucker}}{{Gorenstein} et~al.}{1978}]{Gorenstein78}
{Gorenstein} P.,  {Fabricant} D.,  {Topka} K.,  {Harnden} F.~R. J.,   {Tucker}
  W.~H.,  1978, \mn@doi [ApJ] {10.1086/156421}, \href
  {https://ui.adsabs.harvard.edu/abs/1978ApJ...224..718G} {224, 718}

\bibitem[\protect\citeauthoryear{{Gunn} \& {Gott}}{{Gunn} \&
  {Gott}}{1972}]{gun72}
{Gunn} J.~E.,  {Gott} J.~R.~I.,  1972, ApJ, \href
  {http://adsabs.harvard.edu/abs/1972ApJ...176....1G} {176, 1}

\bibitem[\protect\citeauthoryear{{Heesen}, {Brinks}, {Leroy}, {Heald}, {Braun},
  {Bigiel}  \& {Beck}}{{Heesen} et~al.}{2014}]{heesen14}
{Heesen} V.,  {Brinks} E.,  {Leroy} A.~K.,  {Heald} G.,  {Braun} R.,  {Bigiel}
  F.,   {Beck} R.,  2014, \mn@doi [AJ] {10.1088/0004-6256/147/5/103}, \href
  {https://ui.adsabs.harvard.edu/abs/2014AJ....147..103H} {147, 103}

\bibitem[\protect\citeauthoryear{{Hester} et~al.,}{{Hester}
  et~al.}{2010}]{hester10}
{Hester} J.~A.,  et~al., 2010, \mn@doi [ApJL] {10.1088/2041-8205/716/1/L14},
  \href {https://ui.adsabs.harvard.edu/abs/2010ApJ...716L..14H} {716, L14}

\bibitem[\protect\citeauthoryear{{Holwerda}, {Pirzkal}, {Cox}, {de Blok},
  {Weniger}, {Bouchard}, {Blyth}  \& {van der Heyden}}{{Holwerda}
  et~al.}{2011}]{holwerda11}
{Holwerda} B.~W.,  {Pirzkal} N.,  {Cox} T.~J.,  {de Blok} W.~J.~G.,  {Weniger}
  J.,  {Bouchard} A.,  {Blyth} S.-L.,   {van der Heyden} K.~J.,  2011, \mn@doi
  [MNRAS] {10.1111/j.1365-2966.2011.18940.x}, \href
  {http://adsabs.harvard.edu/abs/2011MNRAS.416.2426H} {416, 2426}

\bibitem[\protect\citeauthoryear{{J{\'a}chym} et~al.,}{{J{\'a}chym}
  et~al.}{2017}]{jachym17}
{J{\'a}chym} P.,  et~al., 2017, \mn@doi [ApJ] {10.3847/1538-4357/aa6af5}, \href
  {http://cdsads.u-strasbg.fr/abs/2017ApJ...839..114J} {839, 114}

\bibitem[\protect\citeauthoryear{{J{\'a}chym} et~al.,}{{J{\'a}chym}
  et~al.}{2019}]{jachym19}
{J{\'a}chym} P.,  et~al., 2019, \mn@doi [ApJ] {10.3847/1538-4357/ab3e6c}, \href
  {https://ui.adsabs.harvard.edu/abs/2019ApJ...883..145J} {883, 145}

\bibitem[\protect\citeauthoryear{{Keenan}, {Davies}, {Taylor}  \&
  {Minchin}}{{Keenan} et~al.}{2016}]{keenan16}
{Keenan} O.~C.,  {Davies} J.~I.,  {Taylor} R.,   {Minchin} R.~F.,  2016,
  \mn@doi [MNRAS] {10.1093/mnras/stv2684}, \href
  {https://ui.adsabs.harvard.edu/abs/2016MNRAS.456..951K} {456, 951}

\bibitem[\protect\citeauthoryear{{Kennicutt}}{{Kennicutt}}{1998}]{kennicut98}
{Kennicutt} Jr. R.~C.,  1998, \mn@doi [ARA\&A]
  {10.1146/annurev.astro.36.1.189}, \href
  {http://cdsads.u-strasbg.fr/abs/1998ARA%26A..36..189K} {36, 189}

\bibitem[\protect\citeauthoryear{{Kleiner} et~al.,}{{Kleiner}
  et~al.}{2021}]{kleiner21}
{Kleiner} D.,  et~al., 2021, \mn@doi [A\&A] {10.1051/0004-6361/202039898},
  \href {https://ui.adsabs.harvard.edu/abs/2021A&A...648A..32K} {648, A32}

\bibitem[\protect\citeauthoryear{{Lagos}, {Demarco}, {Papaderos}, {Telles},
  {Nigoche-Netro}, {Humphrey}, {Roche}  \& {Gomes}}{{Lagos}
  et~al.}{2016}]{lagos16}
{Lagos} P.,  {Demarco} R.,  {Papaderos} P.,  {Telles} E.,  {Nigoche-Netro} A.,
  {Humphrey} A.,  {Roche} N.,   {Gomes} J.~M.,  2016, \mn@doi [MNRAS]
  {10.1093/mnras/stv2702}, \href
  {http://cdsads.u-strasbg.fr/abs/2016MNRAS.456.1549L} {456, 1549}

\bibitem[\protect\citeauthoryear{{Lagos}, {Scott}, {Nigoche-Netro}, {Demarco},
  {Humphrey}  \& {Papaderos}}{{Lagos} et~al.}{2018}]{lagos18}
{Lagos} P.,  {Scott} T.~C.,  {Nigoche-Netro} A.,  {Demarco} R.,  {Humphrey} A.,
    {Papaderos} P.,  2018, \mn@doi [MNRAS] {10.1093/mnras/sty601}, \href
  {http://cdsads.u-strasbg.fr/abs/2018MNRAS.477..392L} {477, 392}

\bibitem[\protect\citeauthoryear{{Lee}, {Kimm}, {Katz}, {Rosdahl}, {Devriendt}
  \& {Slyz}}{{Lee} et~al.}{2020}]{lee20}
{Lee} J.,  {Kimm} T.,  {Katz} H.,  {Rosdahl} J.,  {Devriendt} J.,   {Slyz} A.,
  2020, \mn@doi [ApJ] {10.3847/1538-4357/abc3b8}, \href
  {https://ui.adsabs.harvard.edu/abs/2020ApJ...905...31L} {905, 31}

\bibitem[\protect\citeauthoryear{{Macquart} et~al.,}{{Macquart}
  et~al.}{2020}]{macquart20}
{Macquart} J.~P.,  et~al., 2020, \mn@doi [Nature] {10.1038/s41586-020-2300-2},
  \href {https://ui.adsabs.harvard.edu/abs/2020Natur.581..391M} {581, 391}

\bibitem[\protect\citeauthoryear{{Maybhate}, {Masiero}, {Hibbard}, {Charlton},
  {Palma}, {Knierman}  \& {English}}{{Maybhate} et~al.}{2007}]{maybhate07}
{Maybhate} A.,  {Masiero} J.,  {Hibbard} J.~E.,  {Charlton} J.~C.,  {Palma} C.,
   {Knierman} K.~A.,   {English} J.,  2007, \mn@doi [MNRAS]
  {10.1111/j.1365-2966.2007.12265.x}, \href
  {http://adsabs.harvard.edu/abs/2007MNRAS.381...59M} {381, 59}

\bibitem[\protect\citeauthoryear{{McMullin}, {Waters}, {Schiebel}, {Young}  \&
  {Golap}}{{McMullin} et~al.}{2007}]{mcmull07}
{McMullin} J.~P.,  {Waters} B.,  {Schiebel} D.,  {Young} W.,   {Golap} K.,
  2007, in {Shaw} R.~A.,  {Hill} F.,   {Bell} D.~J.,  eds,  Astronomical
  Society of the Pacific Conference Series Vol. 376, Astronomical Data Analysis
  Software and Systems XVI. p.~127

\bibitem[\protect\citeauthoryear{{Moretti} et~al.,}{{Moretti}
  et~al.}{2020}]{moretti20}
{Moretti} A.,  et~al., 2020, \mn@doi [ApJ] {10.3847/1538-4357/ab616a}, \href
  {https://ui.adsabs.harvard.edu/abs/2020ApJ...889....9M} {889, 9}

\bibitem[\protect\citeauthoryear{{Moss}}{{Moss}}{2006}]{moss06}
{Moss} C.,  2006, \mn@doi [MNRAS] {10.1111/j.1365-2966.2006.11000.x}, \href
  {http://adsabs.harvard.edu/abs/2006MNRAS.373..167M} {373, 167}

\bibitem[\protect\citeauthoryear{{M{\"u}ller} et~al.,}{{M{\"u}ller}
  et~al.}{2021}]{muller21}
{M{\"u}ller} A.,  et~al., 2021, \mn@doi [Nature Astronomy]
  {10.1038/s41550-020-01234-7}, \href
  {https://ui.adsabs.harvard.edu/abs/2021NatAs...5..159M} {5, 159}

\bibitem[\protect\citeauthoryear{{Poggianti} et~al.,}{{Poggianti}
  et~al.}{2017}]{poggiant17}
{Poggianti} B.~M.,  et~al., 2017, \mn@doi [ApJ] {10.3847/1538-4357/aa78ed},
  \href {https://ui.adsabs.harvard.edu/abs/2017ApJ...844...48P} {844, 48}

\bibitem[\protect\citeauthoryear{{Poggianti} et~al.,}{{Poggianti}
  et~al.}{2019}]{poggianti19}
{Poggianti} B.~M.,  et~al., 2019, \mn@doi [MNRAS] {10.1093/mnras/sty2999},
  \href {https://ui.adsabs.harvard.edu/abs/2019MNRAS.482.4466P} {482, 4466}

\bibitem[\protect\citeauthoryear{{Poggianti} et~al.,}{{Poggianti}
  et~al.}{2020}]{poggianti20}
{Poggianti} B.~M.,  et~al., 2020, ArXiv 170310301Y, \href
  {https://ui.adsabs.harvard.edu/abs/2020arXiv200503735P} {p. arXiv:2005.03735}

\bibitem[\protect\citeauthoryear{{Ramatsoku} et~al.,}{{Ramatsoku}
  et~al.}{2019}]{Ramatsoku19}
{Ramatsoku} M.,  et~al., 2019, \mn@doi [MNRAS] {10.1093/mnras/stz1609}, \href
  {https://ui.adsabs.harvard.edu/abs/2019MNRAS.487.4580R} {487, 4580}

\bibitem[\protect\citeauthoryear{{Reynolds} et~al.,}{{Reynolds}
  et~al.}{2021}]{Reynolds21}
{Reynolds} T.~N.,  et~al., 2021, \mn@doi [MNRAS] {10.1093/mnras/stab1371},
  \href {https://ui.adsabs.harvard.edu/abs/2021MNRAS.505.1891R} {505, 1891}

\bibitem[\protect\citeauthoryear{{Robitaille} \& {Bressert}}{{Robitaille} \&
  {Bressert}}{2012}]{Robitaille2012}
{Robitaille} T.,  {Bressert} E.,  2012, {APLpy: Astronomical Plotting Library
  in Python}, Astrophysics Source Code Library (\mn@eprint {ascl} {1208.017})

\bibitem[\protect\citeauthoryear{{Roediger} \& {Br{\"u}ggen}}{{Roediger} \&
  {Br{\"u}ggen}}{2007}]{roed07}
{Roediger} E.,  {Br{\"u}ggen} M.,  2007, \mn@doi [MNRAS]
  {10.1111/j.1365-2966.2007.12241.x}, \href
  {http://adsabs.harvard.edu/abs/2007MNRAS.tmp..765R} {p.~765}

\bibitem[\protect\citeauthoryear{{Scott} et~al.,}{{Scott}
  et~al.}{2010}]{scott10}
{Scott} T.~C.,  et~al., 2010, \mn@doi [MNRAS]
  {10.1111/j.1365-2966.2009.16204.x}, \href
  {http://adsabs.harvard.edu/abs/2010MNRAS.403.1175S} {403, 1175 (Paper I)}

\bibitem[\protect\citeauthoryear{{Scott}, {Cortese}, {Brinks}, {Bravo-Alfaro},
  {Auld}  \& {Minchin}}{{Scott} et~al.}{2012}]{scott12}
{Scott} T.~C.,  {Cortese} L.,  {Brinks} E.,  {Bravo-Alfaro} H.,  {Auld} R.,
  {Minchin} R.,  2012, \mn@doi [MNRAS] {10.1111/j.1745-3933.2011.01169.x},
  \href {http://adsabs.harvard.edu/abs/2012MNRAS.419L..19S} {419, L19 (Paper
  II)}

\bibitem[\protect\citeauthoryear{{Scott}, {Brinks}, {Cortese}, {Boselli}  \&
  {Bravo-Alfaro}}{{Scott} et~al.}{2018}]{scott18}
{Scott} T.~C.,  {Brinks} E.,  {Cortese} L.,  {Boselli} A.,   {Bravo-Alfaro} H.,
   2018, \mn@doi [MNRAS] {10.1093/mnras/sty063}, \href
  {http://cdsads.u-strasbg.fr/abs/2018MNRAS.475.4648S} {475, 4648}

\bibitem[\protect\citeauthoryear{{Serra} et~al.,}{{Serra}
  et~al.}{2013}]{serra13}
{Serra} P.,  et~al., 2013, \mn@doi [MNRAS] {10.1093/mnras/sts033}, \href
  {http://adsabs.harvard.edu/abs/2013MNRAS.428..370S} {428, 370}

\bibitem[\protect\citeauthoryear{{Serra} et~al.,}{{Serra}
  et~al.}{2015}]{serra15}
{Serra} P.,  et~al., 2015, \mn@doi [MNRAS] {10.1093/mnras/stv079}, \href
  {https://ui.adsabs.harvard.edu/abs/2015MNRAS.448.1922S} {448, 1922}

\bibitem[\protect\citeauthoryear{{Smith} et~al.,}{{Smith}
  et~al.}{2010}]{smith10}
{Smith} R.~J.,  et~al., 2010, \mn@doi [MNRAS]
  {10.1111/j.1365-2966.2010.17253.x}, \href
  {https://ui.adsabs.harvard.edu/abs/2010MNRAS.408.1417S} {408, 1417}

\bibitem[\protect\citeauthoryear{{Solanes}, {Manrique},
  {Garc{\'{\i}}a-G{\'o}mez}, {Gonz{\'a}lez-Casado}, {Giovanelli}  \&
  {Haynes}}{{Solanes} et~al.}{2001}]{sola01}
{Solanes} J.~M.,  {Manrique} A.,  {Garc{\'{\i}}a-G{\'o}mez} C.,
  {Gonz{\'a}lez-Casado} G.,  {Giovanelli} R.,   {Haynes} M.~P.,  2001, \mn@doi
  [ApJ] {10.1086/318672}, \href
  {http://adsabs.harvard.edu/abs/2001ApJ...548...97S} {548, 97}

\bibitem[\protect\citeauthoryear{{Speagle}, {Steinhardt}, {Capak}  \&
  {Silverman}}{{Speagle} et~al.}{2014}]{speagle14}
{Speagle} J.~S.,  {Steinhardt} C.~L.,  {Capak} P.~L.,   {Silverman} J.~D.,
  2014, \mn@doi [ApJS] {10.1088/0067-0049/214/2/15}, \href
  {https://ui.adsabs.harvard.edu/abs/2014ApJS..214...15S} {214, 15}

\bibitem[\protect\citeauthoryear{{Spergel} et~al.,}{{Spergel}
  et~al.}{2007}]{sperg07}
{Spergel} D.~N.,  et~al., 2007, \mn@doi [ApJS] {10.1086/513700}, \href
  {http://adsabs.harvard.edu/abs/2007ApJS..170..377S} {170, 377}

\bibitem[\protect\citeauthoryear{{Steinhauser}, {Schindler}  \&
  {Springel}}{{Steinhauser} et~al.}{2016}]{Steinhauser16}
{Steinhauser} D.,  {Schindler} S.,   {Springel} V.,  2016, \mn@doi [A\&A]
  {10.1051/0004-6361/201527705}, \href
  {https://ui.adsabs.harvard.edu/abs/2016A&A...591A..51S} {591, A51}

\bibitem[\protect\citeauthoryear{{Sun}, {Donahue}, {Roediger}, {Nulsen},
  {Voit}, {Sarazin}, {Forman}  \& {Jones}}{{Sun} et~al.}{2010}]{sun10}
{Sun} M.,  {Donahue} M.,  {Roediger} E.,  {Nulsen} P.~E.~J.,  {Voit} G.~M.,
  {Sarazin} C.,  {Forman} W.,   {Jones} C.,  2010, \mn@doi [ApJ]
  {10.1088/0004-637X/708/2/946}, \href
  {http://adsabs.harvard.edu/abs/2010ApJ...708..946S} {708, 946}

\bibitem[\protect\citeauthoryear{{Tamburro}, {Rix}, {Leroy}, {Mac Low},
  {Walter}, {Kennicutt}, {Brinks}  \& {de Blok}}{{Tamburro}
  et~al.}{2009}]{tamburu09}
{Tamburro} D.,  {Rix} H.~W.,  {Leroy} A.~K.,  {Mac Low} M.~M.,  {Walter} F.,
  {Kennicutt} R.~C.,  {Brinks} E.,   {de Blok} W.~J.~G.,  2009, \mn@doi [AJ]
  {10.1088/0004-6256/137/5/4424}, \href
  {https://ui.adsabs.harvard.edu/abs/2009AJ....137.4424T} {137, 4424}

\bibitem[\protect\citeauthoryear{{Tempel}, {Tago}  \& {Liivam{\"a}gi}}{{Tempel}
  et~al.}{2012}]{tempel12}
{Tempel} E.,  {Tago} E.,   {Liivam{\"a}gi} L.~J.,  2012, \mn@doi [A\&A]
  {10.1051/0004-6361/201118687}, \href
  {https://ui.adsabs.harvard.edu/abs/2012A&A...540A.106T} {540, A106}

\bibitem[\protect\citeauthoryear{{Vikhlinin}, {Forman}  \& {Jones}}{{Vikhlinin}
  et~al.}{1999}]{Vikhlinin99}
{Vikhlinin} A.,  {Forman} W.,   {Jones} C.,  1999, \mn@doi [ApJ]
  {10.1086/307876}, \href
  {https://ui.adsabs.harvard.edu/abs/1999ApJ...525...47V} {525, 47}

\bibitem[\protect\citeauthoryear{{Vollmer}, {Balkowski}, {Cayatte}, {van Driel}
   \& {Huchtmeier}}{{Vollmer} et~al.}{2004}]{voll04}
{Vollmer} B.,  {Balkowski} C.,  {Cayatte} V.,  {van Driel} W.,   {Huchtmeier}
  W.,  2004, \mn@doi [A\&A] {10.1051/0004-6361:20034552}, \href
  {http://adsabs.harvard.edu/abs/2004A%26A...419...35V} {419, 35}

\bibitem[\protect\citeauthoryear{{Wang}, {Koribalski}, {Serra}, {van der
  Hulst}, {Roychowdhury}, {Kamphuis}  \& {Chengalur}}{{Wang}
  et~al.}{2016}]{wang16}
{Wang} J.,  {Koribalski} B.~S.,  {Serra} P.,  {van der Hulst} T.,
  {Roychowdhury} S.,  {Kamphuis} P.,   {Chengalur} J.~N.,  2016, \mn@doi
  [MNRAS] {10.1093/mnras/stw1099}, \href
  {http://cdsads.u-strasbg.fr/abs/2016MNRAS.460.2143W} {460, 2143}

\bibitem[\protect\citeauthoryear{{Yagi} et~al.,}{{Yagi} et~al.}{2010}]{yagi10}
{Yagi} M.,  et~al., 2010, \mn@doi [AJ] {10.1088/0004-6256/140/6/1814}, \href
  {https://ui.adsabs.harvard.edu/abs/2010AJ....140.1814Y} {140, 1814}

\bibitem[\protect\citeauthoryear{{Yagi}, {Yoshida}, {Gavazzi}, {Komiyama},
  {Kashikawa}  \& {Okamura}}{{Yagi} et~al.}{2017}]{yagi17}
{Yagi} M.,  {Yoshida} M.,  {Gavazzi} G.,  {Komiyama} Y.,  {Kashikawa} N.,
  {Okamura} S.,  2017, \mn@doi [ApJ] {10.3847/1538-4357/aa68e3}, \href
  {http://esoads.eso.org/abs/2017ApJ...839...65Y} {839, 65}

\bibitem[\protect\citeauthoryear{{Yoshida} et~al.,}{{Yoshida}
  et~al.}{2008}]{yoshida08}
{Yoshida} M.,  et~al., 2008, \mn@doi [ApJ] {10.1086/592430}, \href
  {https://ui.adsabs.harvard.edu/abs/2008ApJ...688..918Y} {688, 918}

\makeatother
\end{thebibliography}
\newpage
\appendix

\onecolumn 
\section{}
\begin{figure*}
\begin{center}
\includegraphics[ angle=0,scale=0.48] {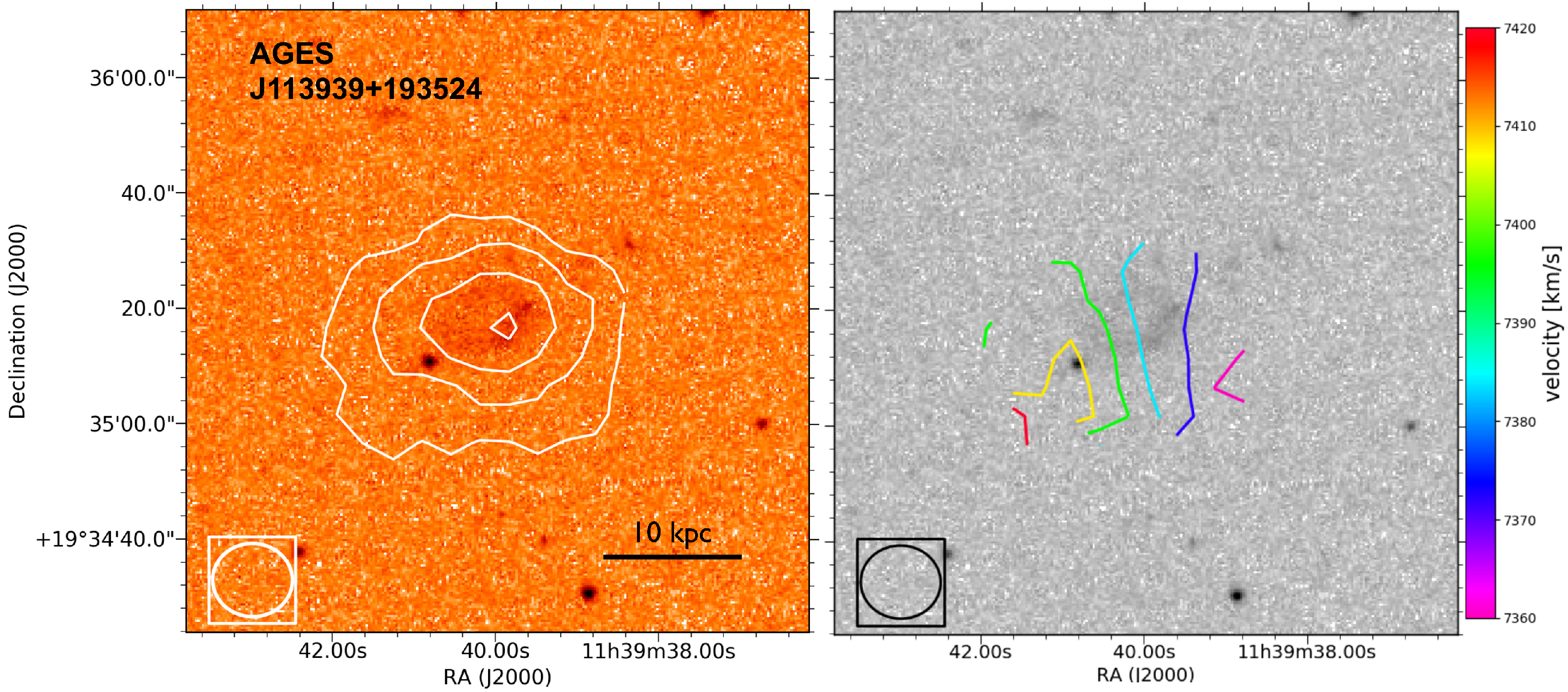}
\vspace{1cm}
\caption{AGES J113939+193524 (SDSS J113939.67+193516.7): {\em left} VLA B/C--configuration \hi\ \textcolor{black}{column} density contours (in white). The contours are at \hi\ column densities of 0.5, 1.6, 3.1, and $4.7 \times 10^{20}$\,atom\,cm$^{-2}$. {\em Right}  \hi\ velocity field  with contours separated by 10 \km.  The background images are SDSS $g$--band. The boxed ellipse at the bottom left of each panel shows the size (13.84  $\times$ 12.69 arcsec) and orientation of  the VLA beam. }
\label{figa}
\end{center}
\end{figure*}

\hi\ was detected in  AGES J113939+193524 (SDSS J113939.67+193516.7) in the FGC\,1287 VLA field of view. SDSS J113939.67+193516.7 is projected $\sim$ 178\,kpc east of FGC\,1287. Figure \ref{figa} shows the  \hi\ \textcolor{black}{column} density  and  velocity field for the galaxy.

\section{}

\begin{figure*}
\begin{center}
\includegraphics[ angle=0,scale=0.48] {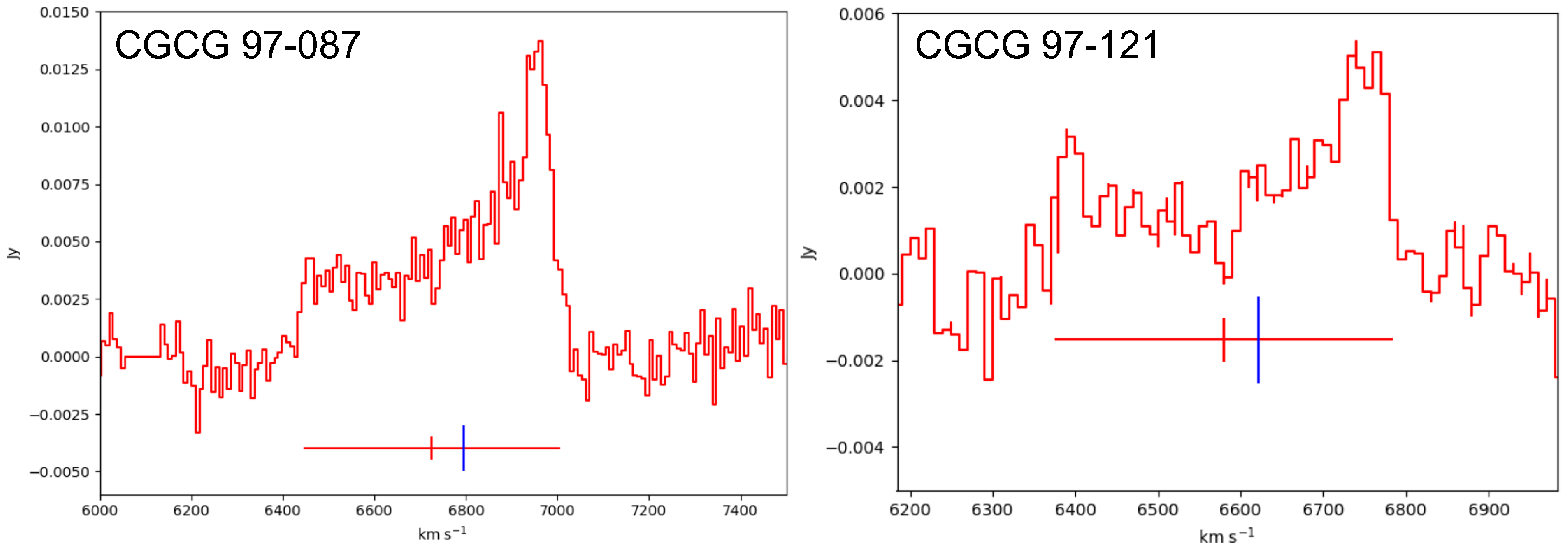}
\vspace{1cm}
\caption{AGES  \hi\ profiles for \cg087 (left) and \cg121 (right). The red horizontal bar at the base of each spectrum delimits  W$_{20}$ with the red vertical bars indicating V$_{HI}$ based on mean W$_{20}$ velocities and the blue vertical bars indicating the flux weighed V$_{HI}$.   }
\label{figb}
\end{center}
\end{figure*}

Figure \ref{figb} shows the AGES \hi\ profiles of \cg087 (left) and \cg121 (right) which are A\,1367 LTGs displaying one--sided \hi\ tails in their VLA integrated maps \citep{scott10,scott18}. The \af\ \hi\  asymmetry measurement for the galaxies are 1.82$\pm$0.09 and 1.92$\pm$0.08 respectively; both are  projected within 550\,kpc of the cluster centre.

\label{lastpage}
\end{document}